\documentclass[aps,reprint,twocolumn,longbibliography,nobalancelastpage,superscriptaddress]{revtex4-1}
\usepackage{amsmath,amssymb,amsfonts}
\usepackage[pdftex]{graphicx}
\usepackage{bm}
\usepackage{dsfont}
\usepackage{color}
\usepackage[colorlinks=True,linkcolor=blue,citecolor=blue,urlcolor=blue]{hyperref}
\usepackage{hypcap}
\newcommand{\x}{\sigma^x}

\newcommand{\z}{\sigma^z}
\newcommand{\upket}{\vert \varepsilon,\uparrow\rangle}
\newcommand{\dnket}{\vert \varepsilon,\downarrow\rangle}
\newcommand{\upbra}{\langle \varepsilon,\uparrow\vert}
\newcommand{\dnbra}{\langle \varepsilon,\downarrow\vert}

\newcommand{\tr}{\mathrm{Tr}}
\newcommand{\dota}{\text{\textbullet}}
\newcommand{\dotb}{\blacktriangle}
\newcommand{\dotc}{\blacktriangledown}
\begin{document}
\title{Nonequilibrium quantum order at infinite temperature: spatiotemporal correlations and their generating functions}

\author{Sthitadhi Roy}
\email{sthitadhi.roy@chem.ox.ac.uk}
\affiliation{Physical and Theoretical Chemistry, Oxford University, South Parks Road, Oxford OX1 3QZ, United Kingdom}
\affiliation{Rudolf Peierls Centre for Theoretical Physics, Clarendon Laboratory, Oxford University, Parks Road, Oxford OX1 3PU, United Kingdom}

\author{Achilleas Lazarides}
\email{acl@pks.mpg.de}
\affiliation{Max-Planck-Institut f{\"u}r Physik komplexer Systeme, N{\"o}thnitzer Stra{\ss}e 38, 01187 Dresden, Germany}

\date{\today}

\begin{abstract}
Localisation-protected quantum order extends the idea of symmetry breaking and order in ground states to individual eigenstates at arbitrary energy. Examples include many-body localised static and $\pi$-spin glasses in Floquet systems. Such order is inherently dynamical and difficult to detect as the order parameter typically varies randomly between different eigenstates, requiring specific superpositions of eigenstates to be targeted by the initial state. We show that two-time correlators overcome this, reflecting the presence or absence of eigenstate order even in fully-mixed, \emph{infinite temperature} states. We show how spatiotemporal correlators are generated by the recently introduced dynamical potentials, demonstrating this explicitly using an Ising and a Floquet $\pi$-spin glass and focusing on features mirroring those of equilibrium statistical mechanics such as bimodal potentials in the symmetry-broken phase.
\end{abstract}

\maketitle

\section{Introduction}

Traditionally, phases of matter and transitions between them have been
a notion restricted to equilibrium and understood largely using
Landau's theory of broken symmetries~\cite{landau2013course}. Very
recently, many-body
localisation~\cite{gornyi2005interacting,basko2006metal,oganesyan2007localisation,znidaric2008many,pal2010many,vosk2015theory,luitz2015many,nandkishore2015many,lev2015absence,abanin2017recent,schreiber2015observation,choi2016exploring}
has led to the introduction of localisation-protected
quantum order~\cite{huse2013localisation} which has ushered in a new
paradigm of order in quantum matter. In phases hosting such order in
static systems, individual many-body eigenstates at arbitrary energy
densities spontaneously break symmetries of the Hamiltonian and
usually exhibit random glassy
order~\cite{huse2013localisation,pekker2014hilbert,kjall2014many,parameswaran2018many}. The
notion of eigenstate phases has also been extended to the class of
periodically-driven, or Floquet, systems, where fundamentally new
phases have been
proposed~\cite{khemani2016phase,keyserlingk2016absolute,moessner2017equilibration}
and
observed~\cite{zhang2017observation,choi2017observation,pal2017rigidity,rovny2018observation}.

Such novel out-of-equilibrium phases raise some fundamental questions,
two of which we address in this work:
\begin{itemize}
\item Firstly, the presence of order in all eigenstates of the system
  naturally suggests that dynamical order parameters be used to
  characterise the phases. However, the eigenstate order is random
  over both space and energy and hence in generic initial states
  having overlap with many eigenstates, the order gets washed out in
  time-dependent expectations of observables.  Therefore probes robust
  to initial conditions, in particular extreme situations like
  infinite temperature ensembles, are of interest.

\item Secondly, as eigenstate phases and transitions are dynamic,
  rather than thermodynamic phenomena, it is not possible to study
  them with the usual tools of statistical mechanics and
  thermodynamics such as effective potentials and free energies. Hence
  a framework for studying the statistical mechanics of such dynamical
  order is naturally an interesting question.
\end{itemize}

We resolve the first of the two issues by showing that
out-of-equilibrium \emph{spatiotemporal correlations} robustly encode
the presence or absence of eigenstate order, remaining a good
diagnostic even with the system in an infinite-temperature state where
its density matrix is proportional to identity. Hence, the
presence of an eigenstate-ordered phase can be diagnosed for
\emph{arbitrary} initial states. This is of practical importance in
cases where coupling to an external environment is significant (such
as in solid-state systems, trapped-ion systems etc.) since the
resulting thermal state will necessarily involve an incoherent mixture
of eigenstates.

The usefulness of spatiotemporal correlations hints towards a possible
direction for addressing the second question. In order to develop a
statistical mechanics-like framework for eigenstate phases, one should
try to construct effective potentials which act as generating
functions for such correlations, much like the free energy in
equilibrium. We do this by extending the dynamical potentials
introduced in a previous work~\cite{roy2017dynamical} to mixed states
such as the infinite temperature states already mentioned. The
connection to spatiotemporal correlations lies in the fact that these
potentials can be recast as probability distributions, moments of which
correspond to various spatiotemporal correlations. These potentials
and hence the probability distributions are found to exhibit
qualitatively different behaviours in different eigenstate ordered
phases.  For example, in a $\mathbb{Z}_2$ symmetric Ising spin-glass,
which spontaneously breaks the $\mathbb{Z}_2$ symmetry in the
spin-glass phase at all energy densities, we find that appropriately
constructed distributions are bimodal whereas in the paramagnet phase,
the same distributions are unimodal with a vanishing width in the thermodynamic limit. The
bimodal distribution guarantees a finite two-point (in space and time)
correlation function for finite systems and hence is a signature for
spontaneous symmetry breaking in the thermodynamic
limit~\cite{kaplan1989order,koma1993symmetry}. The unimodal
distribution with a vanishing width in
the paramagnet phase on the other hand shows the absence of any
long-ranged (in space and time) correlation or order. This is
analogous to double (single)-well free energy potentials in ordered
(disordered) phases in equilibrium statistical mechanics.  The
framework then provides a statistical mechanics-esque way of describing
eigenstate order phases macroscopically which is also robust to
infinite temperature ensembles and hence is expected to work for any
generic initial condition, and constitutes the central result of this
work.

To concretely demonstrate our results, we explicitly construct the
potentials for the two prototypical examples of eigenstate ordered
systems, namely, a $\mathbb{Z}_2$ symmetric disordered Ising chain
hosting a spin glass-paramagnet phase
transition~\cite{huse2013localisation} and its periodically driven
cousin hosting a $\pi$-spin glass/discrete time crystal
phase~\cite{khemani2016phase} exclusive to Floquet systems.

The rest of the paper is organised as follows. In
Sec.~\ref{sec:quantum-order}, we discuss the phenomenology of
eigenstate order in a disordered Ising spin chain and its periodically
driven version, and demonstrate how spatiotemporal correlations at
infinite temperature encode the eigenstate
order. Sec.~\ref{sec:potential} generalises the framework of dynamical
potentials for mixed states, followed by their explicit numerical
constructions and discussions on the results for the two models in
Sec.~\ref{sec:numerical}. In Sec.~\ref{sec:bimodality}, we display
results for a particular operator which leads to bimodal potentials
(and probability distributions) in the eigenstate ordered phases
before finally concluding with a summary and outlook in
Sec.~\ref{sec:conclusion}.

\section{Spatiotemporal correlations and quantum order}
\label{sec:quantum-order}

\subsection{Phenomenology of eigenstate order}
\label{sec:phen-eigenst-order}

\subsubsection{Static}

The paradigmatic system displaying localisation protected quantum
order is the $\mathbb{Z}_2$ symmetric disordered
Ising chain in one dimension with the Hamiltonian~\cite{huse2013localisation}
\begin{equation}
\mathcal{H}_\mathrm{ISG} = \sum_\ell [J_\ell \z_\ell\z_{\ell+1} + J_x\x_\ell\x_{\ell+1} + h_\ell \x_\ell],
\label{eq:hamising}
\end{equation}
where $J_\ell\in[J_z-J,J_z+J]$ and $h_\ell \in [h_x-h,h_x+h]$ denote
random spin-spin interactions and fields respectively.  For
$J\gg J_z,J_x,h$, the model hosts an eigenstate ordered phase with
Ising spin glass order which can be captured by an Edwards-Anderson
order parameter.  In this phase disorder pins the domain walls
spatially leading to random glassy order in the system. As a result,
the $\mathbb{Z}_2$ symmetry is spontaneously broken at all energy
densities (rather than just in the ground states, as in the clean
Ising ferromagnet).

Concretely, since the parity operator $\mathcal{P}=\prod_\ell \x_\ell$
commutes with the Hamiltonian \eqref{eq:hamising}, the eigenstates of
$\mathcal{H}_\mathrm{ISG}$ are eigenstates of
$\mathcal{P}$ simultaneously.  In the spin glass phase the eigenstates of
$\mathcal{H}_\mathrm{ISG}$ are each two-fold degenerate (up to
corrections exponentially small in system size) with each member of
the pair having opposite parity.  We write these states as
$\vert \varepsilon,\pm\rangle$, where
\begin{equation}
\mathcal{H}_\mathrm{ISG}\vert \varepsilon,\pm\rangle=\varepsilon\vert \varepsilon,\pm\rangle
;~~\mathcal{P}\vert \varepsilon,\pm\rangle=\pm\vert \varepsilon,\pm\rangle.
\label{eq:parityeigen}
\end{equation}
Each eigenstate has long-ranged order along the $\sigma^z$ direction:
\begin{equation}
\langle\varepsilon\pm\vert\z_i\z_j\vert\varepsilon,\pm\rangle\neq 0; ~\vert i-j\vert\to \infty.
\end{equation}
The presence of spontaneously symmetry-broken order in the
thermodynamic limit becomes explicit if the eigenstates
$\vert\varepsilon,\pm\rangle$ are expressed in the symmetry broken
basis
\begin{equation}
\vert\varepsilon,\pm\rangle = \frac{1}{\sqrt{2}}[\upket\pm\dnket];~~\upket=\mathcal{P}\dnket.
\label{eq:symmbrokenbasis}
\end{equation}
where the order is evident:
\begin{equation}
\upbra\z_i\upket = -\dnbra\z_i\dnket =\eta_{i,\varepsilon}\neq 0.
\label{eq:symmetrybrokenorder}
\end{equation}
The $\eta_{i,\varepsilon}$ are not smooth functions of $\varepsilon$,
changing randomly between eigenstates close in energy. They also
remain finite with increasing system size.  
Eqs.~\eqref{eq:symmbrokenbasis} and \eqref{eq:symmetrybrokenorder} imply
\begin{equation}
\langle\varepsilon,\pm\vert\z_i\vert \varepsilon,\mp\rangle = \eta_{i,\varepsilon}.
\label{eq:matelparity}
\end{equation}

The model also hosts a paramagnetic phase for $h,h_x\gg J,J_x$, where
the $\eta_{i,\varepsilon}$s defined in
Eq.~\eqref{eq:symmetrybrokenorder} vanish in the thermodynamic limit.

\subsubsection{Floquet}

The presence of $\mathbb{Z}_2$ symmetry and localisation also allows
for a fundamentally different out-of-equilibrium phase, called the
$\pi$-spin glass or ``Discrete Time Crystal'', in the periodically
driven cousin of the model. Note that disorder here is fundamentally important to prevent
heating under the
driving~\cite{dalessio2014long,lazarides2014equilibrium,ponte2015periodically,lazarides2015fate,ponte2015many,bordia2017periodically,reitter2017interaction}.
The discrete time crystalline behaviour of the system shows up in the form of a subharmonic signal. The model is described
by the time-dependent Hamiltonian of unit period
\begin{equation}
  \mathcal{H}_{\pi\mathrm{SG}}(t)  =\begin{cases}
    \sum_\ell [J_\ell \z_\ell\z_{\ell+1} + J_x\x_\ell\x_{\ell+1}]; n\leq t<n+\frac{1}{2}\\
    \sum_\ell [h_\ell \x_\ell + J_x\x_\ell\x_{\ell+1} ]; n+\frac{1}{2}\leq t<n+1
  \end{cases}
  \label{eq:hampisg}
\end{equation}
where $n$ takes integer values. Much of the phenomenology of the
spatial glassy order of the Ising spin glass carries over to the
eigenstates of the Floquet unitary operator
$U_F = e^{-i\int_0^1 dt~\mathcal{H}_{\pi\mathrm{SG}}(t)}$ which again
come in pairs:
\begin{equation}
  U_F \vert\varepsilon,\pm\rangle = \pm e^{-i\varepsilon}\vert\varepsilon,\pm\rangle,
\label{eq:parityeigengfloq}
\end{equation}
where the quantity $\varepsilon$ plays the role of energy, is only
defined modulo the driving frequency $2\pi$ and is
therefore called the quasienergy~\cite{eckardt2017atomic}. The difference from the static case
is that the two parity-related eigenstates
$\vert\varepsilon,\pm\rangle$ are no longer degenerate in quasienergy
but rather separated by half the driving frequency, that is, by $\pi$.
Switching again to the symmetry broken basis, the extra $\pi$ phase
between the two eigenstates leads to the stroboscopic evolution
\begin{equation}
  U_F^n\vert \varepsilon,\uparrow/\downarrow\rangle = e^{-in\varepsilon}\vert \varepsilon,\downarrow/\uparrow\rangle
\end{equation}
and hence
$\langle\varepsilon,{\uparrow/\downarrow}\vert\z_i(n)\vert
\varepsilon,{\uparrow/\downarrow}\rangle =
\pm(-1)^n\eta_{i,\varepsilon}$ such that the stroboscopic
response has a period twice that of the Hamiltonian
\eqref{eq:hampisg}, reducing the discrete time translation symmetry of
the Hamiltonian by a factor of two. This is a direct consequence of
the pairing in the spectrum of $U_F$.

In conclusion, the $\pi$-spin glass displays temporal order with a frequency
which is half the frequency of driving in addition to the spatial
order.

\subsection{Spatiotemporal correlations at infinite temperature} 
The eigenstate order in both $0$ and $\pi$-spin glass is random, both
spatially and between eigenstates. While realistic schemes for
preparing the system such that the order is visible (essentially,
preparing it in a superposition dominated by two eigenstates forming
one of the pairs) are possible, an obvious question is whether there
is some signature of this order in high-temperature mixed
states. Focusing on the extreme limit of the infinite-temperature
state where the density matrix is proportional to identity, we will
show that the answer to this is affirmative, provided that the
appropriate operators defining the potential (the $\mathcal{O}$ of
Sec.~\ref{sec:potential}) are selected.

Let us begin by showing what the difficulty is and how it is
circumvented in the $0$-SG. The order, indicated by finite
$\eta_{i,\varepsilon}$ in Eq.~\eqref{eq:symmetrybrokenorder}, is
random in magnitude and sign over space and eigenstates.

As a result,
dynamical expectation values of the macroscopic version of the
observable, $M_z=\sum_\ell \z_l$, average out to zero:
$\langle M_z(t)\rangle_\infty \to 0$ as $t\to\infty$ where
the $\infty$ in the subscript denotes the expectation value over an
infinite temperature state,
$\langle M_z(t)\rangle_\infty = \mathrm{Tr}[M_z(t)\rho_\infty]$ with
$\rho_\infty
=\mathbb{I}_{\mathrm{dim}[\mathcal{H}]}/\mathrm{dim}[\mathcal{H}]$.
Nevertheless, as we will now show, two-time correlations remain finite
even at infinite temperatures capturing the presence of spatiotemporal
order. This applies to both the static Ising spin glass and the
$\pi$-spin glass: Consider the two time correlator of the longitudinal
magnetisation
\begin{equation}
  \mathcal{C}_{M_z}(t_1,t_2) = \tr[M_z(t_1)M_z(t_2)\rho_\infty],
\label{eq:twotimecorr1}
\end{equation}
which for the Ising spin glass using Eq.~\eqref{eq:parityeigen} can be expressed as 
\begin{widetext}
\begin{equation}
  \mathcal{C}_{M_z}(t_1,t_2) = \frac{1}{\mathrm{dim}[\mathcal{H}]}\sum_{\varepsilon,\varepsilon^\prime,\varepsilon^{\prime \prime}}\sum_{\dota,\dotb,\dotc=\pm}
  \left[\langle \varepsilon,\dota\vert M_z\vert\varepsilon^\prime,\dotb\rangle
    \langle \varepsilon^\prime,\dotb\vert\varepsilon^{\prime\prime},\dotc\rangle
    \langle \varepsilon^{\prime\prime},\dotc\vert M_z\vert \varepsilon,\dota\rangle~
    e^{it_1(\varepsilon-\varepsilon^\prime)}e^{it_2(\varepsilon^{\prime\prime}-\varepsilon)}\right].
\label{eq:twotimecorr2}
\end{equation}
\end{widetext}
The time-averaged two-time correlator is then defined as
\begin{equation}
\bar{\mathcal{C}}_{M_z}(t) = \frac{1}{t^2}\int_0^t dt_1 \int_0^t dt_2~ \mathcal{C}_{M_z}(t_1,t_2),
\label{eq:averagedtwotime}
\end{equation}
which in the limit of $t\to\infty$ can be expressed using Eq.~\eqref{eq:matelparity} as
\begin{eqnarray}
  \bar{\mathcal{C}}_{M_z}(t\to\infty) &=& \frac{2}{\mathrm{dim}[\mathcal{H}]}\sum_{\varepsilon}\sum_{i,j=1}^{L}\eta_{i,\varepsilon}\eta_{j,\varepsilon}\nonumber\\
                                      &\approx& \frac{2}{\mathrm{dim}[\mathcal{H}]}\sum_{\varepsilon}\sum_{i}^{L}\eta_{i,\varepsilon}^2\sim O(L),
\label{eq:twotimecorrelavg}
\end{eqnarray}
where terms of the form
$\sum_{i\neq j}\eta_{i,\varepsilon}\eta_{j,\varepsilon}$ vanish in the
thermodynamic limit as $\eta_{i,\varepsilon}$ can take random
signs. On the other hand, since the average magnitudes of
$\eta_{i,\varepsilon}$ do not vanish in the spin glass phase in the
thermodynamic limit, terms of the form in
Eq.~\eqref{eq:twotimecorrelavg} survive and yield a $O(L)$
contribution to the two-time correlator. Contrary to the spin glass
phase, in the paramagnetic phase $\eta_{i,\varepsilon}$ vanishes in
magnitude in the thermodynamic limit and consequently so does the
two-time correlator. Hence, the question that whether the dynamical
order in the spin glass phase survives infinite temperature is
answered in the affirmative and the two-time correlation of the
macroscopic longitudinal magnetisation carries the information of the
order.

A similar analysis shows the presence of the temporal order in
the case of the $\pi$-spin glass, where the stroboscopic
two-time correlator can be expressed as
\begin{widetext}
\begin{align}
  \mathcal{C}_{M_z}(n_1,n_2) = \frac{1}{\mathrm{dim}[\mathcal{H}]}\sum_{\varepsilon,\varepsilon^\prime,\varepsilon^{\prime \prime}}\sum_{\dotb,\dotc=\pm}
  e^{i\varepsilon(n_1-n_2)}[&\langle \varepsilon,+\vert M_z\vert\varepsilon^\prime,\dotb\rangle
                              \langle \varepsilon^\prime,\dotb\vert U_F^{n_1-n_2}\vert \varepsilon^{\prime\prime},\dotc\rangle
                              \langle \varepsilon^{\prime\prime},\dotc\vert M_z\vert \varepsilon,+\rangle+\nonumber\\
                            &\langle \varepsilon,-\vert M_z\vert\varepsilon^\prime,\dotb\rangle
                              \langle \varepsilon^\prime,\dotb\vert U_F^{n_1-n_2}\vert \varepsilon^{\prime\prime},\dotc\rangle
                              \langle \varepsilon^{\prime\prime},\dotc\vert M_z\vert \varepsilon,-\rangle(-1)^{n_1+n_2}].
\end{align}
\end{widetext}
Using Eqs.~\eqref{eq:matelparity} and \eqref{eq:parityeigengfloq} in
the long-time limit this can be expressed as
\begin{equation}
\mathcal{C}_{M_z}(n_1,n_2) \approx (-1)^{n_1+n_2}\frac{2}{\mathrm{dim}[\mathcal{H}]}\sum_{\varepsilon}\sum_{i}^{L}\eta_{i,\varepsilon}^2,
\label{eq:twotimecorrelfloq}
\end{equation}
which again is $O(L)$ and hence non-vanishing in the thermodynamic limit but more crucially, oscillates with a period twice the stroboscopic time and hence reflects the discrete time crystalline order in addition to the spatial spin glass order.

\section{Dynamical potentials for mixed states \label{sec:potential}}
Having established that temporal correlation functions are the key
towards exposing the eigenstate order at infinite temperature, we
generalise the framework of dynamical potentials introduced in
\cite{roy2017dynamical} for pure states to mixed states.

Consider a system governed by a Hamiltonian $\mathcal{H}(t)$ and let
the observable of interest be $\mathcal{O}$. If the initial state of
the system at $t=0$ is described by the density matrix $\rho$, then
one can define a functional
\begin{equation}
\mathcal{Z}_t[s] = \tr[\mathcal{T}e^{-i\int_0^t dt^\prime\mathcal{H}^{(s)}(t^\prime)}\rho \tilde{\mathcal{T}}e^{i\int_0^t dt^\prime\mathcal{H}^{(s)\dagger}(t^\prime)}]
\label{eq:zst}
\end{equation}
where 
\begin{equation}\mathcal{H}^{(s)}(t) = \mathcal{H}(t)-is(t)\mathcal{O}/2,
\label{eq:hs}
\end{equation} and $\mathcal{T}(\tilde{\mathcal{T}})$ denote (anti-)time orderings.
$\mathcal{Z}_t[s]$ acts as the moment generating functional for the correlations of $\mathcal{O}$ as 
\begin{align}
  &\frac{\delta \mathcal{Z}}{\delta s(t)}\bigg\vert_{s=0}=\tr[\mathcal{O}(t)\rho],\label{eq:first-moment}\\
  &\frac{\delta^2 \mathcal{Z}}{\delta s(t_1)\delta s(t_2)}\bigg\vert_{s=0}=\frac{1}{2}\left[\tr[\mathcal{O}(t_1)\mathcal{O}(t_2)\rho]+t_1\leftrightarrow t_2\right],\label{eq:second-moment}
\end{align}
and so on. In the case of a constant $s$, $\mathcal{Z}(s,t)$ takes on
the integrated form of the temporal correlations as
\begin{align}
  &\frac{\partial \mathcal{Z}(s,t)}{\partial s}\bigg\vert_{s=0}=\int _0^t dt^\prime~\tr[\mathcal{O}(t^\prime)\rho],\\
  &\frac{\partial^2 \mathcal{Z}(s,t)}{\partial s^2}\bigg\vert_{s=0}=\int_0^tdt_1\int_0^tdt_2\tr[\mathcal{O}(t_1)\mathcal{O}(t_2)\rho],
\end{align}
and so on. One can define $\Theta(s,t) = L\theta(s,t)$ as the corresponding cumulant generating function via
\begin{equation}
  \mathcal{Z}(s,t) = e^{-\Theta(s,t)}.
  \label{eq:theta}
\end{equation}

As in Ref.~\cite{roy2017dynamical}, the moment generating function can be used to calculate
the a probability distribution whose moments encode the temporal
correlations of various orders. Expressing $\mathcal{Z}(s,t)$ as
\begin{equation}
\mathcal{Z}(s,t) = \int dm~e^{-smL}P(m,t,L),
\end{equation}
where $P(m,t,L)\sim e^{-\Phi(m,t,L)}$ with $\Phi(m,t,L)=L\phi(m,t,L)$, one can simply calculate $P$ from $\phi$ via a Legendre transform 
\begin{equation}
  \phi(m,t) = -\max_s[s m - \theta(s,t)].
\label{eq:phi}
\end{equation}

The potentials $\theta$ and $\phi$ effectively contain all the
information of the dynamics of the system at infinite temperature in
the form of temporal correlations.  As shown in the following
sections, they exhibit qualitatively different behaviours in different
phases, and hence provide for a macroscopic characterisation of such
eigenstate phases.

\section{Numerical results \label{sec:numerical}}
As spatiotemporal correlations can show eigenstate order at infinite
temperature and the dynamical potentials provide a general framework
for studying them, we present pertinent numerical results for the
Ising spin glass \eqref{eq:hamising} as well as for the periodically
driven model \eqref{eq:hampisg}. We choose
$\mathcal{O}=M_z=\sum_\ell\z_l$, the total longitudinal magnetisation.

\subsection{Ising spin glass \label{sec:isgnumerics}}

\begin{figure}
\includegraphics[width=\columnwidth]{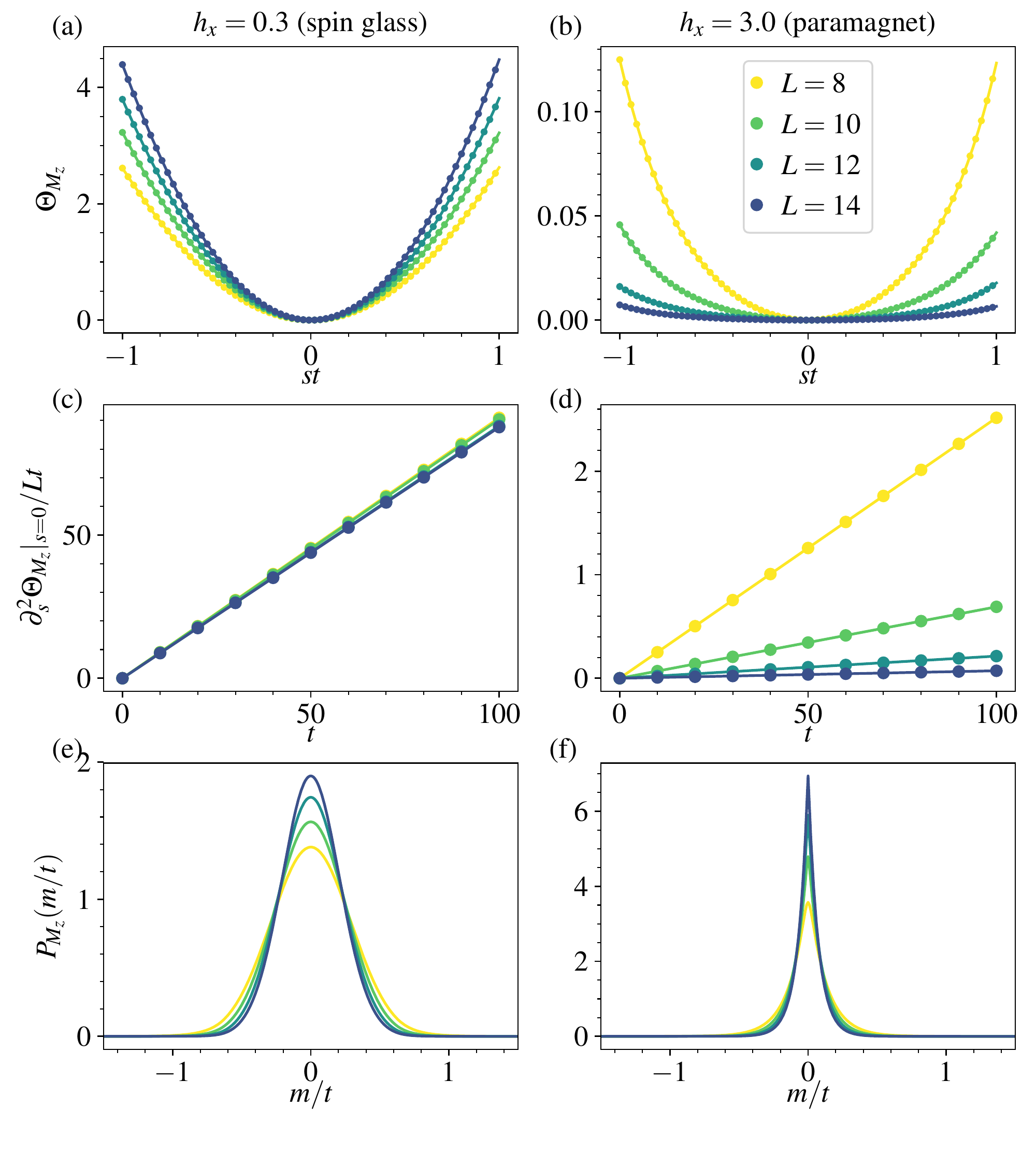}
\caption{The dynamical potential $\Theta_{M_z}$ as function of $st$
  for different system sizes, $L$, in the spin glass (a) and
  paramagnet (b) phases respectively. Note that they have opposite
  trends with increasing $L$ with regard to their curvature at
  $s=0$. The second derivative of $\Theta_{M_z}$ with respect to $s$
  at $s=0$ as function of $t$ scaled with $L$ and $t$ suggests that
  $\Theta_{M_z}\sim s^2t^2L$ in the spin glass (c) and vanishes in the
  thermodynamic limit in the paramagnet (d)phase. The distribution
  $P_{M_z}(m/t)$ shows a Gaussian and exponential behaviour in the
  spin glass (e) and paramagnet (f) phases respectively. Other
  parameters are $J_z=0$, $J=5$, $J_x=0.3$, and $h=0.3$.  }
\label{fig:isg_mag}
\end{figure}

We numerically compute the potentials corresponding to the Ising spin
glass for two different parameter values expected to be in the spin
glass and paramagnetic phases. The potentials are calculated with $s$
constant in time so we can get time-integrated temporal correlation
functions of the form in Eq.~\eqref{eq:averagedtwotime}. The results
for the disorder averaged potential $\Theta_{M_z}$ are shown in
Fig.~\ref{fig:isg_mag}(a) and (b) and they clearly show different
behaviours in the two phases. In the spin glass phase, the curvature
at $s=0$ increases with increasing $L$, whereas in the paramagnet
phase the curvature decreases with increasing $L$. This is shown more
clearly in Fig.~\ref{fig:isg_mag}(c) and (d) where we we plot the
second derivative of $\Theta_{M_z}$ with respect to $s$ at $s=0$
scaled with $L$ and $t$. The collapse of the data and linear behaviour
of $\partial_s^2\Theta\vert_{s=0}/Lt$ with $t$ in the spin glass phase
suggests a scaling of the form
$\partial_s^2\Theta_{M_z}\vert_{s=0}\sim L t^2$, reflecting the
presence of spin glass order and in agreement with the scaling
predicted from the phenomenology in
Eq.~\eqref{eq:twotimecorrelavg}. On the other hand, in the paramagnet
phase, not just the second but all derivatives of $\Theta$ with
respect to $s$ at $s=0$ seem to vanish in the thermodynamic limit (see
Fig.~\ref{fig:isg_mag}(b) and (d)). Hence, the dynamical potential
$\Theta_{M_z}$ shows that the eigenstate order can be probed via a
macroscopic observable at infinite temperature.

As mentioned in Sec.~\ref{sec:potential}, we can also construct the
distribution $P$, moments of which yield temporal correlations of all
orders. In Fig.~\ref{fig:isg_mag}(e) and (f) we show $P$ in both spin
glass and paramagnetic phases using Eq.~\eqref{eq:phi}. They exhibit
qualitatively different behaviour: while in the spin glass $P$ has a
Gaussian form with standard deviation proportional to $1/\sqrt{L}$, in the
paramagnetic phase the distribution has an exponential form. The
origin of the Gaussian in the spin glass phase can be understood
easily as the leading contribution to $\Theta_{M_z}$ in the
thermodynamic limit is $\sim s^2t^2L$ and hence the leading
contribution to $\Phi\sim m^2L/t^2$.  On the other hand, in the
paramagnet phase, since all derivatives of $\Theta$ with respect to
$s$ at $s=0$ decrease with increasing $L$, the distribution decays
exponentially away from the mean.

Note that the first derivative of $\Theta$ with respect to $s$ at
$s=0$ vanishes in the both the phases, indicating that the expectation
value of $M_z$ as a function of $t$ cannot capture the difference
between the spin glass and the paramagnet phases, thus highlighting
the importance of multiple-time correlations. This is also manifested
in the fact that the mean of the distribution $P$ is zero in both the
phases.

\subsection{Periodically driven Ising spin glass \label{sec:pisgnumerics}}
\begin{figure}
\includegraphics[width=\columnwidth]{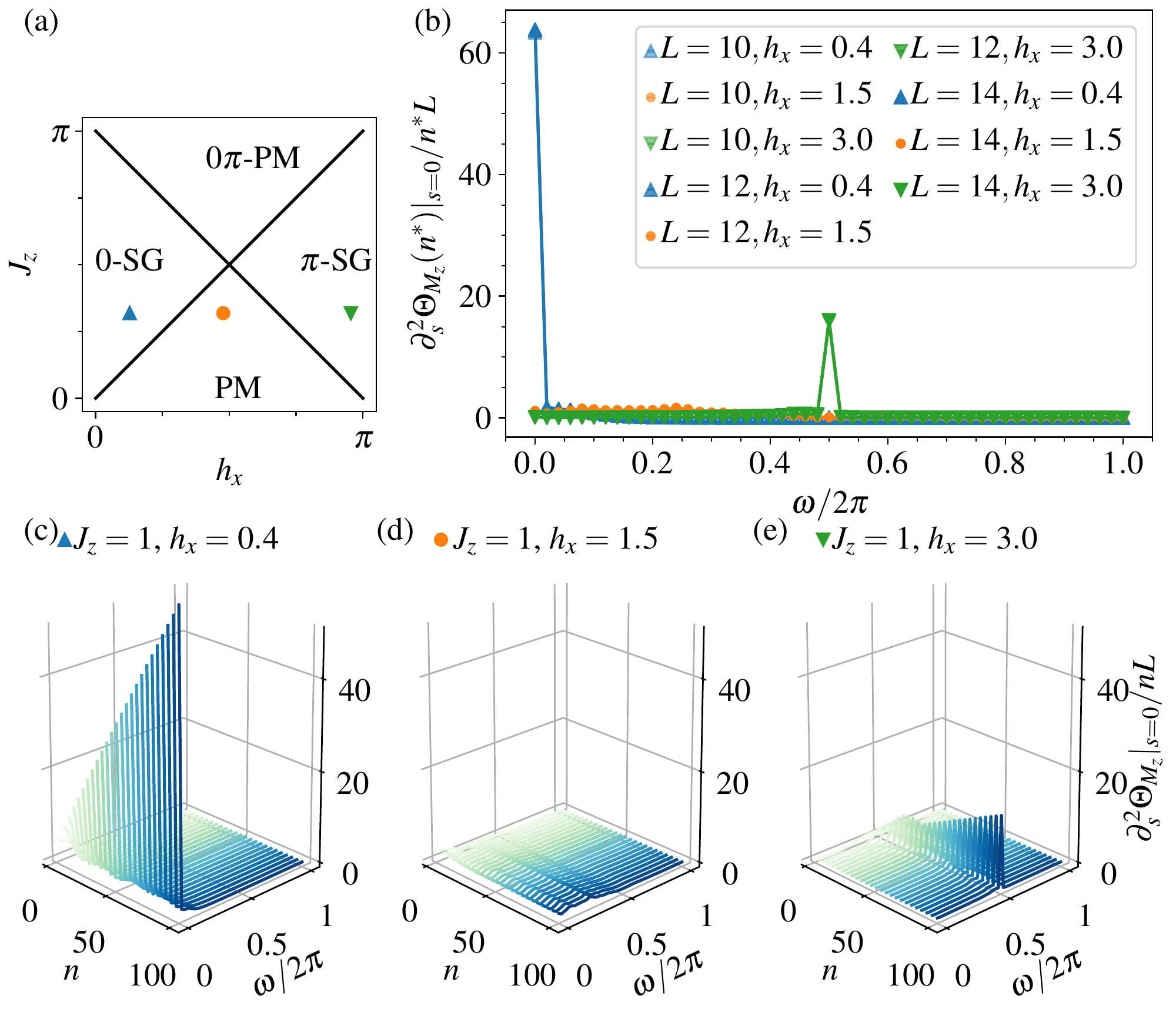}
\caption{(a) The phase diagram of the model in Eq.~\eqref{eq:hampisg} for $J_x=0$ and the markers shows the parameter values of $J_z$ and $h_x$ corresponding to which, the numerical results are shown in rest of the panels. (b) $\partial_s^2\Theta_{M_z}(n^\ast)\vert_{s=0}/n^\ast L$ as function of the frequency $\omega$ at fixed time $n^\ast=100$ for the three different $(J_z,h_x)$ values marked in (a) for different $L$. The collapse of the data for different $L$ suggests that the response survives in the thermodynamic limit. (c)-(e) The behaviour of $\partial_s^2\Theta_{M_z}\vert_{s=0}/n L$ as function of $\omega$ and $n$ together shows that the integrated two-time correlator grows quadratically with time, if probed at $\omega=0$ and $\pi$ in the $0$- and $\pi$-spin glass phases respectively. Other parameters are $J=5$, $J_x=0.1$, $h=0.3$, and $L=14$ for (c)-(e).}
\label{fig:dtc}
\end{figure}

The periodically driven cousin of the disordered Ising model
\eqref{eq:hampisg} in the limit of $J_x=0$ has an exactly known phase
diagram shown in Fig.~\ref{fig:dtc}(a) and the phases are known to be
stable to the presence of interactions provided the system stays in
the Floquet-many body localised phase.  Since the model is driven
periodically in time (with period $1$) and thus can have non-trivial
temporal behaviour, we use a time-dependent $s$ of the form
\begin{equation}
s(t) = s\cos(\omega t),
\label{eq:st}
\end{equation}
and hence the potentials now have an additional parameter, namely the
frequency, $\omega$, of the source field.  With this form of $s(t)$,
we calculate the $\Theta_{M_z}$ for three parameter values, corresponding to the
0-spin glass, paramagnet, and the $\pi$-spin glass, as shown by the
markers in Fig.~\ref{fig:dtc}(a).

Since the $0$-spin glass has a phenomenology identical to the Ising spin glass discussed in Sec.~\ref{sec:isgnumerics}, $\partial_s^2\Theta\vert_{s=0}\sim s^2n^2L$ ($n$ being the stroboscopic time) for $\omega\to 0$. This can be inferred from the data for $h_x=0.4$ from Fig.~\ref{fig:dtc}(b) and (c).

More interestingly, in the $\pi$-spin glass, a similar response is
present for $\omega=\pi$, \textit{i.e.} at half the frequency of the
periodic drive and hence shows a time crystalline response
(Fig.~\ref{fig:dtc}(b) and (e)). The response at $\omega=\pi$ can be
understood quite simply from Eq.~\eqref{eq:twotimecorrelfloq} as the
two-time correlator $\mathcal{C}_{M_z}(n_1,n_2)$has a subharmonic
response in both, $n_1$ and $n_2$. However, the more remarkable aspect
is that the information of the subharmonic response in the two-time
correlator survives even in an infinite temperature ensemble despite
the spatial order being random over space and eigenstates.

Finally, in the paramagnet phase, since there is no spatial order anyway, there is no response at any frequency as can be seen in Fig.~\ref{fig:dtc}(b) and (d).

\section{Bimodality and non-analytic potentials}
\label{sec:bimodality}

In Sec.~\ref{sec:potential}, we saw that the probability distribution
for the chosen operator (analogous to the exponential of a
thermodynamic potential) is obtained from $\mathcal{Z}_t(s)$ of
Eq.~(\ref{eq:zst}) (we fix $s$ to be constant for simplicity) by a
Legendre transform. In a phase with a broken $\mathbb{Z}_2$ symmetry
one might expect that the probability distribution for an
appropriately chosen operator will be bimodal, analogous to a free
energy landscape with a double well structure or the effective
potential in a field theory~\cite{mussardo2010statistical}. As we show
later, the operator
\begin{equation}
  G_z = \sum_{i\neq j}\z_i\z_j,
\label{eq:Gz}
\end{equation}
is such an appropriate operator.

A practical issue is that the probability distribution obtained after
a naive Legendre transform cannot be multimodal as the transform
preserves convexity. In addition, we will see that the potential
$\Theta$ constructed for the operator $G_z$ appears to be
superextensive in the system size $L$. We now show that these two
issues are both resolved by an appropriate splitting of $\mathcal{Z}$
inspired by the analysis of Ref.~\cite{touchette2010methods} for
calculating non-concave entropies in classical systems.

\subsection{A toy model as a limiting case \label{sec:toy}}

To show how to resolve these two issues, we focus on the classical toy
model defined by the limit $J_x=h_\ell=0=J_z$ and $J=1$ in
Eq.~(\ref{eq:hamising}) and the eigenstates of which are adiabatically
connected to those in the spin-glass phase of Eq.~\eqref{eq:hamising} and the 0-SG phase of Sec.~\ref{sec:quantum-order}.
It turns out that a careful consideration of how the limits of
$s\to 0$ and $L\to\infty$ (see for instance
Eqs.~(\ref{eq:first-moment}) and (\ref{eq:second-moment})) is
necessary, and the correct treatment gives a
non-analytic potential and a bimodal distribution. Our analysis
follows the construction of non-concave entropies in
Ref.~\cite{touchette2010methods}.

In the infinite bond disorder limit ($J_x=h_\ell=0=J_z$ and $J=1$), the energy eigenstates 
are superpositions of pairs of product states which are $\mathbb{Z}_2$
flipped partners (in the $\sigma^z$ basis) of each other. In this limit, $G_z$ commutes with the Hamiltonian so that the energy
eigenstates are also eigenstates of $G_z$.  Labelling the eigenstates
with $\varepsilon$ and $\pm$ analogously to
Sec.~\ref{sec:phen-eigenst-order} and additionally by the difference
between the number of up and down spins in the state, $d$, the
eigenvalues of $G_z$ depends only upon $d$:
\begin{equation}
  G_z\vert \varepsilon,\pm,d\rangle = \overline{G_z(d)}\vert \varepsilon,\pm,d\rangle
\end{equation}
where $\overline{G_z(d)} = (d^2-L)/2$, and there are
$N_d = \binom{L}{(L-d)/2}$ such states in the Hilbert space. The
moment generating function \emph{for a finite system size $L$} is then
\begin{equation*}
  \mathcal{Z}=\frac{1}{2^L}\sum_{d}N_d e^{-st\overline{G_z(d)}}.
\end{equation*}
Calculating $\mathcal{Z}$ and $\Theta$ for a finite system reveals a problem: one can show that in this infinite bond disorder limit,
\begin{eqnarray}
\partial_s\Theta_{G_z}\vert_{s=0} &=& 0, \\
\partial^2_s\Theta_{G_z}\vert_{s=0} &=& t^2(L^2-L)/2 \approx t^2L^2/2
\end{eqnarray}
and hence the leading order, in $s$, term in $\Theta_{G_z}\approx s^2t^2L^2/2$ is superextensive.
This is unphysical as $\Theta$ is purportedly analogous to a thermodynamic potential for the our-of-equilibrium system and hence should be extensive. The issue is resolved by the same
procedure that allows for the $\phi$ (see Eq.~(\ref{eq:phi})) to be
non-convex. We now outline this procedure.

In our calculation the thermodynamic limit $L\rightarrow\infty$ must
be taken. Since $\overline{G_z(d)}$ may take both positive and
negative values, $-L/2 \leq\overline{G_z(d)}\leq L(L-1)/2$, we split
up $\mathcal{Z}$
\begin{align}
  \mathcal{Z}&=\mathcal{Z}^+ + \mathcal{Z}^-\nonumber\\
             &=\frac{1}{2^L}\left[\sum_{\vert d\vert\leq\lfloor\sqrt{L}\rfloor}N_de^{-st\overline{G_z(d)}}+\sum_{\lfloor\sqrt{L}\rfloor<\vert d\vert<L}N_de^{-st\overline{G_z(d)}}\right],
\end{align}
where the first term consists of the eigenstates for which
$\overline{G_z(d)}>0$ while the second 
$\overline{G_z(d)}<0$. The limit $L\rightarrow\infty$ now picks out
one of the two terms depending on the sign of $s$, with
$\mathrm{sgn}(s)=\pm 1$ picking out $\mathcal{Z}^\pm$,
respectively. That is, \emph{the generating function $\mathcal{Z}$ is
  non-analytic in the thermodynamic limit}, with 
\begin{equation}
  \label{eq:limits}
  \Theta_{G_z}(s,t)\sim
  \begin{cases}
    \Theta_{G_z}^+(s,t) = \log\mathcal{Z}^+, & s > 0 \\
    \Theta_{G_z}^-(s,t) = \log\mathcal{Z}^-, & s<0.
    \end{cases}
\end{equation}
The potential is to be Legendre transformed as described in
Ref.~\cite{touchette2010methods}: If $\Phi^\pm(m)$ is the Legendre
transform of $\Theta^\pm(s)$ (as in Eq.~(\ref{eq:phi})) then the
correct $\Phi(m)=\max\left(\Phi^+(m),\Phi^-(m)\right)$ might be
non-concave, as is the case for a bimodal probability distribution.
It is also easy to show that $\Theta^\pm\sim\pm\vert s\vert tL$ and
hence the potential is extensive.

Thus, in practice, splitting the $\mathcal{Z}$ as in
Eq.~(\ref{eq:zexpanded}) and taking the thermodynamic limit
\begin{itemize}
\item leads to a non-analytic $\Theta$ which in turns leads to a
  bimodal probability distribution,
\item leads to extensive potentials,
\end{itemize}
and hence resolves the two issues raised at the beginning of this section.
We now apply this to our quantum model.

\subsection{Numerical results for the Ising spin glass}

We numerically compute the potentials with the operator \eqref{eq:Gz} for the Ising spin glass
and show that the physics discussed in the previous subsection holds away from the toy limit of infinite bond disorder as well.

\begin{figure}
\includegraphics[width=\columnwidth]{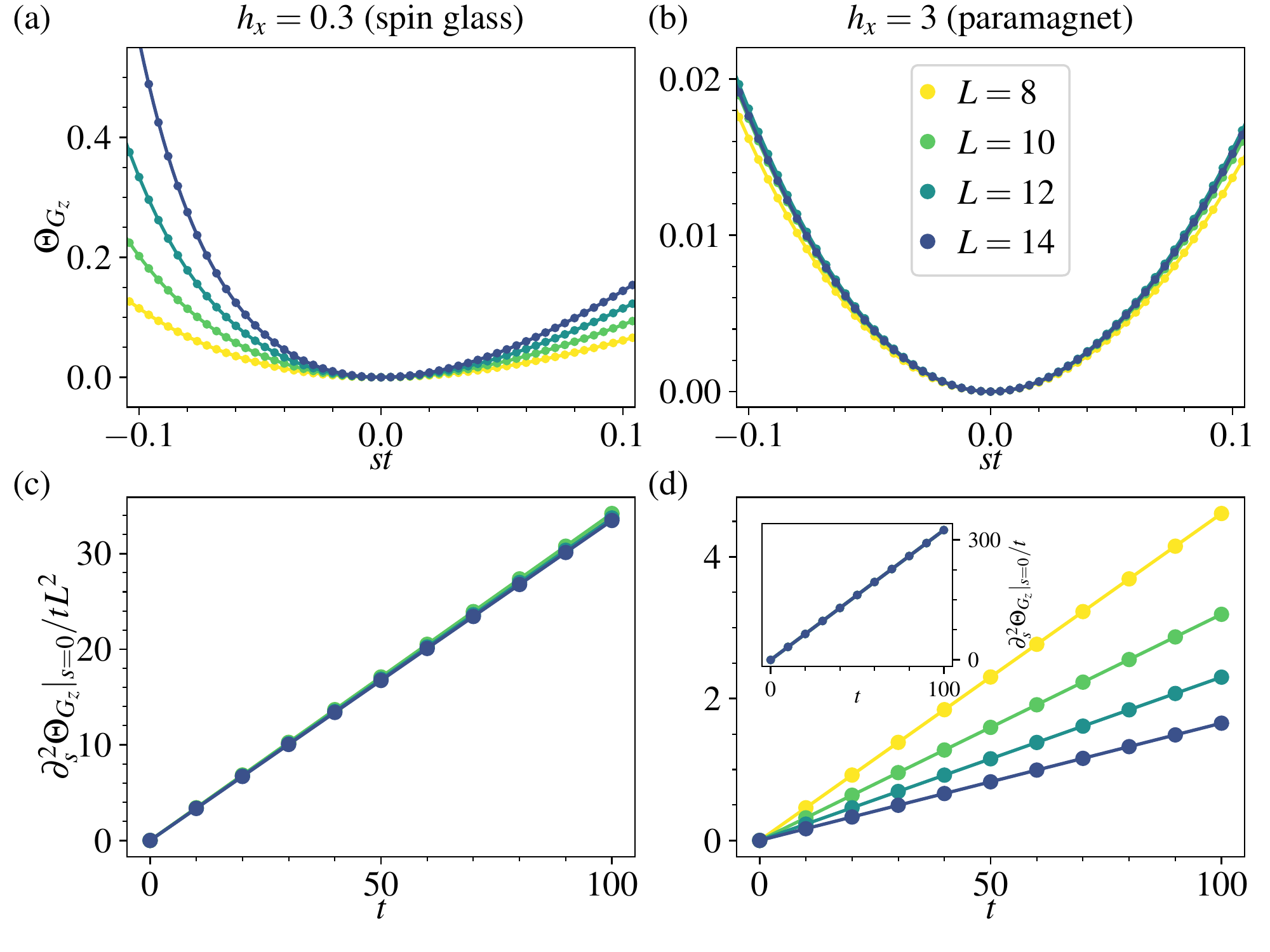}
\caption{The potentials $\Theta_{G_z}$ as a function of $st$ in the spin glass (a) and the paramagnet (b) phases respectively for various $L$. The behaviour of the corresponding $\partial_s^2\Theta_{G_z}\vert_{s=0}^/tL^2$ suggests a form $\Theta_{G_z}\sim s^2t^2L^2$ in the spin glass (c) and $\sim s^2t^2L^0$ in the paramagnet phase (d) (see inset for collapse). Other parameters are $J_z=0$, $J=5$, $J_x=0.3$, and $h=0.3$. }
\label{fig:theta_isg_corr}
\end{figure}

The results for the potential $\Theta_{G_z}$ in the spin glass and
paramagnet phases are shown in Fig.~\ref{fig:theta_isg_corr}~(a) and
(b) respectively. While the curvature of $\Theta_{G_z}$ at $s=0$ seems
to grow with increasing $L$ in the spin glass, it does not seem to
depend on the system size in the paramagnet phase.  To clarify this,
we explicitly look at the behaviour of
$\partial_s^2\Theta_{G_z}\vert_{s=0}$ in
Fig.~\ref{fig:theta_isg_corr}~(c) and (d).

Remarkably, in the spin glass the collapse of
$\partial_s^2\Theta_{G_z}\vert_{s=0}/L^2$ for various system sizes
suggests that the leading contribution to $\Theta_{G_z}$ is of the
form $\sim s^2t^2L^2$. As discussed in Sec.~\ref{sec:toy}, this might seem alarming as $\mathcal{Z}$ plays
the role analogous the partition function and $\Theta$, the total free
energy, which in this case seems to be superextensive in $L$.

The way to resolve apparently nonphysical result that
$\Theta_{G_z}\sim s^2t^2L^2$ for the full spin glass problem is then
that, as discussed in Sec.~\ref{sec:toy}, the limit of $L\to\infty$ should precede $s\to0$, as is common in
the treatment of problems with spontaneously broken symmetry.  As it is
impossible to take the limits in that order in the numerical treatment
of a finite system, we separate the sum defining $\mathcal{Z}$ into
branches by analogy to the example in Sec.~\ref{sec:toy}, as follows.

\begin{figure}
\includegraphics[width=\columnwidth]{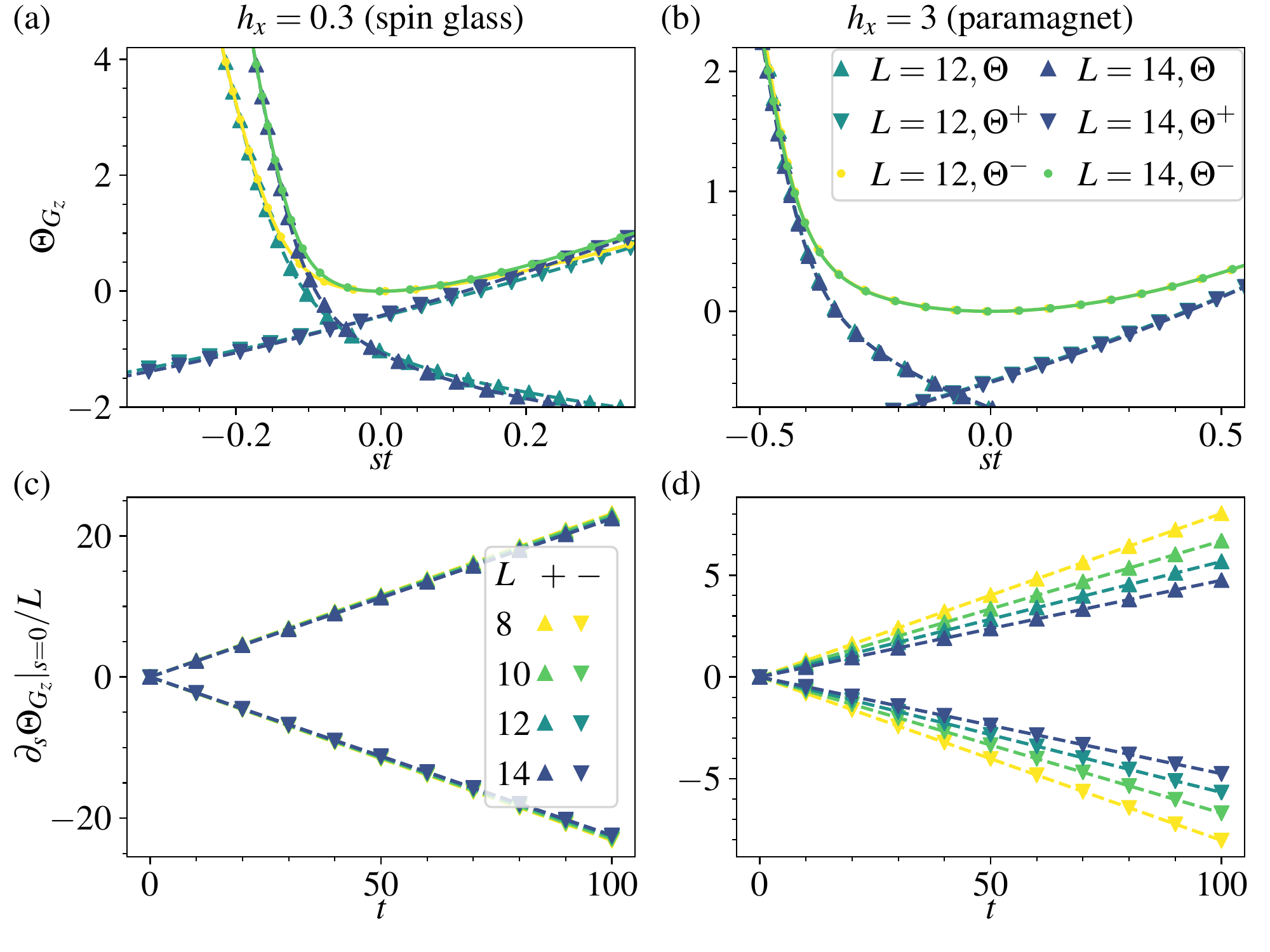}
\caption{The branches $\Theta_{G_z}^\pm$ for the spin glass and
  paramagnet phases are superposed on $\Theta$ for two different
  system sizes for the spin glass (a) and the paramagnet (b) phases
  respectively. The first derivative of the branches with respect to
  $s$ at $s=0$ as function of $t$ and various system sizes suggests
  that their leading behaviour is of the form
  $\Theta_{G_z}^\pm\sim\pm stL$ in the spin glass (c) and
  $\sim stL^{-1}$ in the paramagnet (d) phase.}
\label{fig:theta_isg_corr_branches}
\end{figure}

Consider $\{\vert\alpha\rangle\}$ to be complete set of basis
states such that $\mathcal{Z}$ can be expressed as
\begin{equation}
  \mathcal{Z} = \sum_{\alpha,\beta,\gamma}\langle\alpha\vert e^{-i\mathcal{H}^{(s)}t}\vert\beta\rangle\langle\beta\vert\rho\vert\gamma\rangle\langle\gamma\vert e^{i\mathcal{H}^{(s)\dagger}t}\vert\alpha\rangle.
\label{eq:zexpanded}
\end{equation}
The branches are then defined by restricting the summation in
Eq.~\eqref{eq:zexpanded} such that
$\langle\alpha\vert G_z\vert\beta\rangle-\langle\gamma\vert
G_z\vert\alpha\rangle\gtrless 0$. The results for the branches
computed numerically are shown in
Fig.~\ref{fig:theta_isg_corr_branches}(a) and (b). The behaviour of
the first derivative of $\Theta^\pm$ with respect to $s$ at $s=0$ in
the spin glass phase shown in
Fig.~\ref{fig:theta_isg_corr_branches}(c) suggests a leading behaviour
of the form $\Theta_{G_z}^\pm\sim \pm stL$ which is now perfectly
consistent with the effective free energy being extensive in
$L$.

Thus the correct form of the potential $\Theta_{G_z}$ as the
thermodynamic limit is approached is non-analytic in the vicinity of
$s=0$, yet analytic branches can be constructed each of which is
extensive in $L$.  In the paramagnet phase, since the expectation of
$G_z$ vanishes in the thermodynamic limit, the derivatives of the
potential with respect to $s$ decay with increasing system size and
hence the construction of branches is not necessary.

\begin{figure}
\includegraphics[width=\columnwidth]{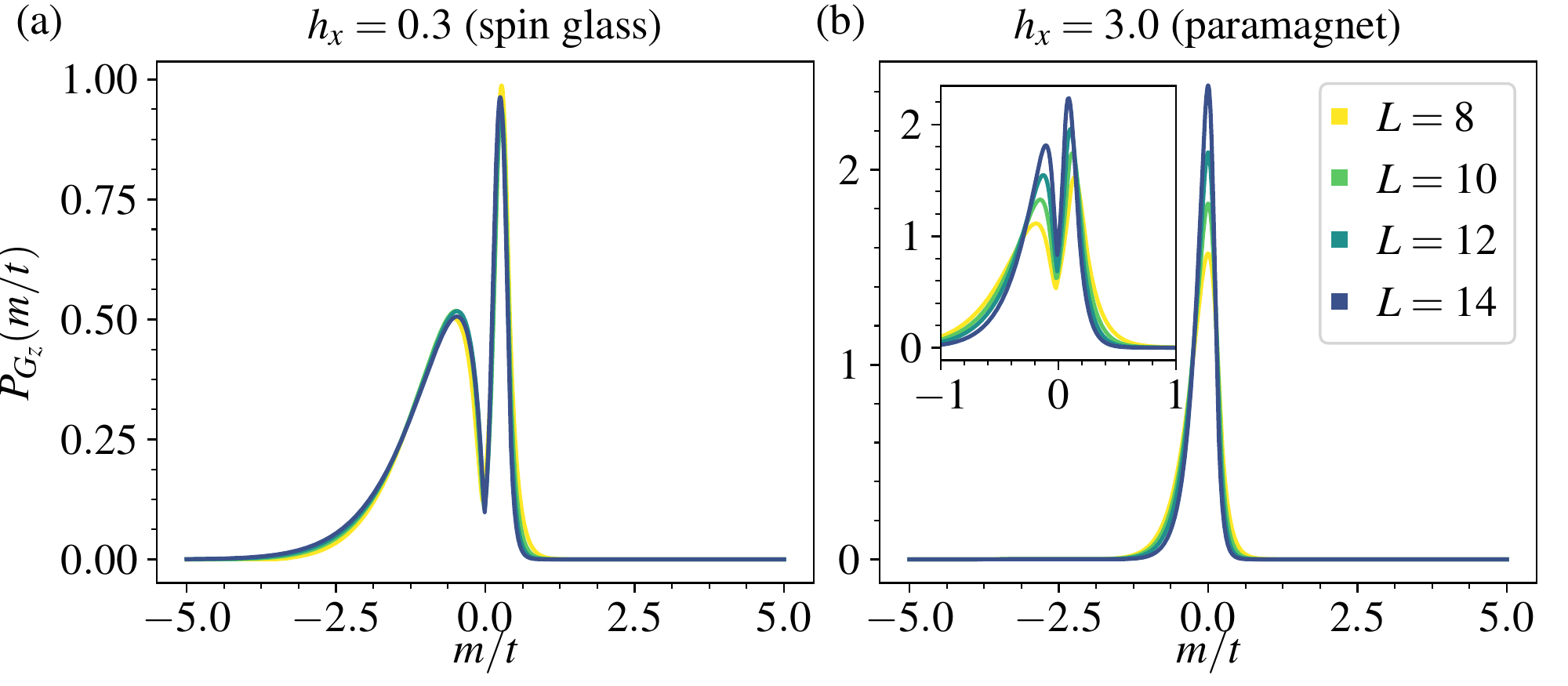}
\caption{The distribution $P_{G_z}$ showing a bimodal nature in the spin glass phase (a) which seems persistent with increasing $L$ and a unimodal distribution in the paramagnet phase (b). The inset shows that if the branch construction is nevertheless used in the paramagnetic phase, one again obtains a bimodal distribution but crucial the bimodality systematically goes away with increasing $L$ eventually converging to the unimodal distribution in the thermodynamic limit.}
\label{fig:P_isg_corr}
\end{figure}
Analogous to the construction of non-concave
entropies~\cite{touchette2010methods}, a non-analytic $\Theta$ would
suggest the presence of a non-concave and hence multimodal
distribution $P$.  The probability distributions are calculated from
the branches individually as $P^\pm\sim e^{-L\phi^\pm}$ with
$\phi^\pm(m,t)=-\max_s[sm-\theta^\pm(s,t)]$ and the overall
probability distribution is reconstructed as
$P(m,t) = max[P^+(m,t),P^-(m)]/\mathcal{N}$, where $\mathcal{N}$ is
overall normalisation factor. The distributions so obtained are shown
in Fig.~\ref{fig:P_isg_corr}. In the spin glass phase the distribution
is bimodal and crucially, there is no indication of the bimodality
systematically going away with increasing system size. We therefore
conclude that the distribution remains bimodal in the thermodynamic
limit. In the paramagnet phase, the branch construction is not used
and the distribution is an unimodal one. Importantly, if one still
does the branch construction in the ferromagnetic phase, a bimodal
distribution is obtained for a finite system but crucially the
bimodality systematically goes away with increasing $L$ (see inset of Fig.~\ref{fig:P_isg_corr}(b)). This
establishes that in the thermodynamic limit, the true distribution is
unimodal.

In conclusion, dynamical potentials for appropriate operators
containing information about their temporal correlations show
qualitatively different behaviours in systems with and without
eigenstate order to the extent that they have different modalities,
even at infinite temperature.

We note in closing that the operator $G_z$ can also be used similarly
for the $\pi$-spin glass except it would show the bimodal distribution
when probed at a frequency $\omega=\pi$.

\section{Conclusion \label{sec:conclusion}}
In conclusion, we have shown that appropriate spatiotemporal
correlations can reflect eigenstate order and symmetry breaking at
arbitrary energy densities in infinite temperature ensembles. A
probe which is completely robust to initial conditions is particularly
advantageous in systems open to the environment which drives the system
to a mixed state involving many eigenstates, washing out eigenstate
order~\cite{medvedyeva2016influence,fischer2016dynamics,levi2016robustness,lazarides2017fate}.
It could also provide a way to get around the issue of eigenstate
order getting washed out due to heating in experimental quantum
simulators like cold-atom or ion-trap systems.

We then generalise the framework of Ref.~\cite{roy2017dynamical} to
mixed states and show that the dynamical potentials in infinite
temperature ensembles show qualitatively different behaviour in
different eigenstate phases. In fact, for appropriate observables, they can
also show a bimodal nature in one phase and unimodal in another.

The numerical treatment presented here is restricted to finite system
sizes thus preventing us from exploring the critical points with
enough accuracy. The next step in this direction is developing
approximate analytical techniques, allowing access to eigenstate
criticality. A natural path towards this could be recasting the moment
generating functional so as to be put on the Keldysh contour. Another
direction is extending this approach to open systems, in which the
time evolution of interest is itself non-unitary.

\begin{acknowledgments}
  We would like to thank R.~Moessner and M.~Heyl for useful
  discussions and collaboration on earlier, related work. This work is in part
  supported by EPSRC Grant No. EP/N01930X/1.
\end{acknowledgments}
\bibliography{refs}

\begin{thebibliography}{41}%
\makeatletter
\providecommand \@ifxundefined [1]{%
 \@ifx{#1\undefined}
}%
\providecommand \@ifnum [1]{%
 \ifnum #1\expandafter \@firstoftwo
 \else \expandafter \@secondoftwo
 \fi
}%
\providecommand \@ifx [1]{%
 \ifx #1\expandafter \@firstoftwo
 \else \expandafter \@secondoftwo
 \fi
}%
\providecommand \natexlab [1]{#1}%
\providecommand \enquote  [1]{``#1''}%
\providecommand \bibnamefont  [1]{#1}%
\providecommand \bibfnamefont [1]{#1}%
\providecommand \citenamefont [1]{#1}%
\providecommand \href@noop [0]{\@secondoftwo}%
\providecommand \href [0]{\begingroup \@sanitize@url \@href}%
\providecommand \@href[1]{\@@startlink{#1}\@@href}%
\providecommand \@@href[1]{\endgroup#1\@@endlink}%
\providecommand \@sanitize@url [0]{\catcode `\\12\catcode `\$12\catcode
  `\&12\catcode `\#12\catcode `\^12\catcode `\_12\catcode `\%12\relax}%
\providecommand \@@startlink[1]{}%
\providecommand \@@endlink[0]{}%
\providecommand \url  [0]{\begingroup\@sanitize@url \@url }%
\providecommand \@url [1]{\endgroup\@href {#1}{\urlprefix }}%
\providecommand \urlprefix  [0]{URL }%
\providecommand \Eprint [0]{\href }%
\providecommand \doibase [0]{http://dx.doi.org/}%
\providecommand \selectlanguage [0]{\@gobble}%
\providecommand \bibinfo  [0]{\@secondoftwo}%
\providecommand \bibfield  [0]{\@secondoftwo}%
\providecommand \translation [1]{[#1]}%
\providecommand \BibitemOpen [0]{}%
\providecommand \bibitemStop [0]{}%
\providecommand \bibitemNoStop [0]{.\EOS\space}%
\providecommand \EOS [0]{\spacefactor3000\relax}%
\providecommand \BibitemShut  [1]{\csname bibitem#1\endcsname}%
\let\auto@bib@innerbib\@empty
\bibitem [{\citenamefont {Landau}\ and\ \citenamefont
  {Lifshitz}(1980)}]{landau2013course}%
  \BibitemOpen
  \bibfield  {author} {\bibinfo {author} {\bibfnamefont {L.~D.}\ \bibnamefont
  {Landau}}\ and\ \bibinfo {author} {\bibfnamefont {E.~M.}\ \bibnamefont
  {Lifshitz}},\ }\href {\doibase
  https://doi.org/10.1016/B978-0-08-057046-4.50002-6} {\emph {\bibinfo {title}
  {Statistical Physics (Course on theoretical physics: Vol. 5)}}},\ \bibinfo
  {edition} {3rd}\ ed.\ (\bibinfo  {publisher} {Butterworth-Heinemann},\
  \bibinfo {address} {Oxford},\ \bibinfo {year} {1980})\BibitemShut {NoStop}%
\bibitem [{\citenamefont {Gornyi}\ \emph {et~al.}(2005)\citenamefont {Gornyi},
  \citenamefont {Mirlin},\ and\ \citenamefont
  {Polyakov}}]{gornyi2005interacting}%
  \BibitemOpen
  \bibfield  {author} {\bibinfo {author} {\bibfnamefont {I.~V.}\ \bibnamefont
  {Gornyi}}, \bibinfo {author} {\bibfnamefont {A.~D.}\ \bibnamefont {Mirlin}},
  \ and\ \bibinfo {author} {\bibfnamefont {D.~G.}\ \bibnamefont {Polyakov}},\
  }\bibfield  {title} {\enquote {\bibinfo {title} {Interacting electrons in
  disordered wires: Anderson localization and low-${T}$ transport},}\ }\href
  {\doibase 10.1103/PhysRevLett.95.206603} {\bibfield  {journal} {\bibinfo
  {journal} {Phys. Rev. Lett.}\ }\textbf {\bibinfo {volume} {95}},\ \bibinfo
  {pages} {206603} (\bibinfo {year} {2005})}\BibitemShut {NoStop}%
\bibitem [{\citenamefont {Basko}\ \emph {et~al.}(2006)\citenamefont {Basko},
  \citenamefont {Aleiner},\ and\ \citenamefont {Altshuler}}]{basko2006metal}%
  \BibitemOpen
  \bibfield  {author} {\bibinfo {author} {\bibfnamefont {D.~M.}\ \bibnamefont
  {Basko}}, \bibinfo {author} {\bibfnamefont {I.~L.}\ \bibnamefont {Aleiner}},
  \ and\ \bibinfo {author} {\bibfnamefont {B.~L.}\ \bibnamefont {Altshuler}},\
  }\bibfield  {title} {\enquote {\bibinfo {title} {Metal--insulator transition
  in a weakly interacting many-electron system with localized single-particle
  states},}\ }\href
  {http://www.sciencedirect.com/science/article/pii/S0003491605002630}
  {\bibfield  {journal} {\bibinfo  {journal} {Annals of {P}hysics}\ }\textbf
  {\bibinfo {volume} {321}},\ \bibinfo {pages} {1126} (\bibinfo {year}
  {2006})}\BibitemShut {NoStop}%
\bibitem [{\citenamefont {Oganesyan}\ and\ \citenamefont
  {Huse}(2007)}]{oganesyan2007localisation}%
  \BibitemOpen
  \bibfield  {author} {\bibinfo {author} {\bibfnamefont {V.}~\bibnamefont
  {Oganesyan}}\ and\ \bibinfo {author} {\bibfnamefont {D.~A.}\ \bibnamefont
  {Huse}},\ }\bibfield  {title} {\enquote {\bibinfo {title} {Localization of
  interacting fermions at high temperature},}\ }\href {\doibase
  10.1103/PhysRevB.75.155111} {\bibfield  {journal} {\bibinfo  {journal} {Phys.
  Rev. B}\ }\textbf {\bibinfo {volume} {75}},\ \bibinfo {pages} {155111}
  (\bibinfo {year} {2007})}\BibitemShut {NoStop}%
\bibitem [{\citenamefont {\ifmmode \check{Z}\else
  \v{Z}\fi{}nidari\ifmmode~\check{c}\else \v{c}\fi{}}\ \emph
  {et~al.}(2008)\citenamefont {\ifmmode \check{Z}\else
  \v{Z}\fi{}nidari\ifmmode~\check{c}\else \v{c}\fi{}}, \citenamefont {Prosen},\
  and\ \citenamefont {Prelov\ifmmode~\check{s}\else
  \v{s}\fi{}ek}}]{znidaric2008many}%
  \BibitemOpen
  \bibfield  {author} {\bibinfo {author} {\bibfnamefont {M.}~\bibnamefont
  {\ifmmode \check{Z}\else \v{Z}\fi{}nidari\ifmmode~\check{c}\else
  \v{c}\fi{}}}, \bibinfo {author} {\bibfnamefont {T.}~\bibnamefont {Prosen}}, \
  and\ \bibinfo {author} {\bibfnamefont {P.}~\bibnamefont
  {Prelov\ifmmode~\check{s}\else \v{s}\fi{}ek}},\ }\bibfield  {title} {\enquote
  {\bibinfo {title} {Many-body localization in the {H}eisenberg {XXZ} magnet in
  a random field},}\ }\href {\doibase 10.1103/PhysRevB.77.064426} {\bibfield
  {journal} {\bibinfo  {journal} {Phys. Rev. B}\ }\textbf {\bibinfo {volume}
  {77}},\ \bibinfo {pages} {064426} (\bibinfo {year} {2008})}\BibitemShut
  {NoStop}%
\bibitem [{\citenamefont {Pal}\ and\ \citenamefont {Huse}(2010)}]{pal2010many}%
  \BibitemOpen
  \bibfield  {author} {\bibinfo {author} {\bibfnamefont {A.}~\bibnamefont
  {Pal}}\ and\ \bibinfo {author} {\bibfnamefont {D.~A.}\ \bibnamefont {Huse}},\
  }\bibfield  {title} {\enquote {\bibinfo {title} {Many-body localization phase
  transition},}\ }\href {\doibase 10.1103/PhysRevB.82.174411} {\bibfield
  {journal} {\bibinfo  {journal} {Phys. Rev. B}\ }\textbf {\bibinfo {volume}
  {82}},\ \bibinfo {pages} {174411} (\bibinfo {year} {2010})}\BibitemShut
  {NoStop}%
\bibitem [{\citenamefont {Vosk}\ \emph {et~al.}(2015)\citenamefont {Vosk},
  \citenamefont {Huse},\ and\ \citenamefont {Altman}}]{vosk2015theory}%
  \BibitemOpen
  \bibfield  {author} {\bibinfo {author} {\bibfnamefont {R.}~\bibnamefont
  {Vosk}}, \bibinfo {author} {\bibfnamefont {D.~A.}\ \bibnamefont {Huse}}, \
  and\ \bibinfo {author} {\bibfnamefont {E.}~\bibnamefont {Altman}},\
  }\bibfield  {title} {\enquote {\bibinfo {title} {Theory of the many-body
  localization transition in one-dimensional systems},}\ }\href {\doibase
  10.1103/PhysRevX.5.031032} {\bibfield  {journal} {\bibinfo  {journal} {Phys.
  Rev. X}\ }\textbf {\bibinfo {volume} {5}},\ \bibinfo {pages} {031032}
  (\bibinfo {year} {2015})}\BibitemShut {NoStop}%
\bibitem [{\citenamefont {Luitz}\ \emph {et~al.}(2015)\citenamefont {Luitz},
  \citenamefont {Laflorencie},\ and\ \citenamefont {Alet}}]{luitz2015many}%
  \BibitemOpen
  \bibfield  {author} {\bibinfo {author} {\bibfnamefont {D.~J.}\ \bibnamefont
  {Luitz}}, \bibinfo {author} {\bibfnamefont {N.}~\bibnamefont {Laflorencie}},
  \ and\ \bibinfo {author} {\bibfnamefont {F.}~\bibnamefont {Alet}},\
  }\bibfield  {title} {\enquote {\bibinfo {title} {Many-body localization edge
  in the random-field {H}eisenberg chain},}\ }\href {\doibase
  10.1103/PhysRevB.91.081103} {\bibfield  {journal} {\bibinfo  {journal} {Phys.
  Rev. B}\ }\textbf {\bibinfo {volume} {91}},\ \bibinfo {pages} {081103}
  (\bibinfo {year} {2015})}\BibitemShut {NoStop}%
\bibitem [{\citenamefont {Nandkishore}\ and\ \citenamefont
  {Huse}(2015)}]{nandkishore2015many}%
  \BibitemOpen
  \bibfield  {author} {\bibinfo {author} {\bibfnamefont {R.}~\bibnamefont
  {Nandkishore}}\ and\ \bibinfo {author} {\bibfnamefont {D.~A.}\ \bibnamefont
  {Huse}},\ }\bibfield  {title} {\enquote {\bibinfo {title} {Many-body
  localization and thermalization in quantum statistical mechanics},}\ }\href
  {http://www.annualreviews.org/doi/10.1146/annurev-conmatphys-031214-014726}
  {\bibfield  {journal} {\bibinfo  {journal} {Annu. Rev. Condens. Matter
  Phys.}\ }\textbf {\bibinfo {volume} {6}},\ \bibinfo {pages} {15} (\bibinfo
  {year} {2015})}\BibitemShut {NoStop}%
\bibitem [{\citenamefont {Bar~Lev}\ \emph {et~al.}(2015)\citenamefont
  {Bar~Lev}, \citenamefont {Cohen},\ and\ \citenamefont
  {Reichman}}]{lev2015absence}%
  \BibitemOpen
  \bibfield  {author} {\bibinfo {author} {\bibfnamefont {Y.}~\bibnamefont
  {Bar~Lev}}, \bibinfo {author} {\bibfnamefont {G.}~\bibnamefont {Cohen}}, \
  and\ \bibinfo {author} {\bibfnamefont {D.~R.}\ \bibnamefont {Reichman}},\
  }\bibfield  {title} {\enquote {\bibinfo {title} {Absence of diffusion in an
  interacting system of spinless fermions on a one-dimensional disordered
  lattice},}\ }\href {\doibase 10.1103/PhysRevLett.114.100601} {\bibfield
  {journal} {\bibinfo  {journal} {Phys. Rev. Lett.}\ }\textbf {\bibinfo
  {volume} {114}},\ \bibinfo {pages} {100601} (\bibinfo {year}
  {2015})}\BibitemShut {NoStop}%
\bibitem [{\citenamefont {Abanin}\ and\ \citenamefont
  {Papi{\'c}}(2017)}]{abanin2017recent}%
  \BibitemOpen
  \bibfield  {author} {\bibinfo {author} {\bibfnamefont {D.~A.}\ \bibnamefont
  {Abanin}}\ and\ \bibinfo {author} {\bibfnamefont {Z.}~\bibnamefont
  {Papi{\'c}}},\ }\bibfield  {title} {\enquote {\bibinfo {title} {Recent
  progress in many-body localization},}\ }\href
  {http://dx.doi.org/10.1002/andp.201700169} {\bibfield  {journal} {\bibinfo
  {journal} {Annalen der Physik}\ }\textbf {\bibinfo {volume} {529}},\ \bibinfo
  {pages} {1700169} (\bibinfo {year} {2017})}\BibitemShut {NoStop}%
\bibitem [{\citenamefont {Schreiber}\ \emph {et~al.}(2015)\citenamefont
  {Schreiber}, \citenamefont {Hodgman}, \citenamefont {Bordia}, \citenamefont
  {L{\"u}schen}, \citenamefont {Fischer}, \citenamefont {Vosk}, \citenamefont
  {Altman}, \citenamefont {Schneider},\ and\ \citenamefont
  {Bloch}}]{schreiber2015observation}%
  \BibitemOpen
  \bibfield  {author} {\bibinfo {author} {\bibfnamefont {M.}~\bibnamefont
  {Schreiber}}, \bibinfo {author} {\bibfnamefont {S.~S.}\ \bibnamefont
  {Hodgman}}, \bibinfo {author} {\bibfnamefont {P.}~\bibnamefont {Bordia}},
  \bibinfo {author} {\bibfnamefont {H.~P.}\ \bibnamefont {L{\"u}schen}},
  \bibinfo {author} {\bibfnamefont {M.~H}\ \bibnamefont {Fischer}}, \bibinfo
  {author} {\bibfnamefont {R.}~\bibnamefont {Vosk}}, \bibinfo {author}
  {\bibfnamefont {E.}~\bibnamefont {Altman}}, \bibinfo {author} {\bibfnamefont
  {U.}~\bibnamefont {Schneider}}, \ and\ \bibinfo {author} {\bibfnamefont
  {I.}~\bibnamefont {Bloch}},\ }\bibfield  {title} {\enquote {\bibinfo {title}
  {Observation of many-body localization of interacting fermions in a
  quasirandom optical lattice},}\ }\href
  {http://science.sciencemag.org/content/349/6250/842} {\bibfield  {journal}
  {\bibinfo  {journal} {Science}\ }\textbf {\bibinfo {volume} {349}},\ \bibinfo
  {pages} {842} (\bibinfo {year} {2015})}\BibitemShut {NoStop}%
\bibitem [{\citenamefont {Choi}\ \emph {et~al.}(2016)\citenamefont {Choi},
  \citenamefont {Hild}, \citenamefont {Zeiher}, \citenamefont {Schau{\ss}},
  \citenamefont {Rubio-Abadal}, \citenamefont {Yefsah}, \citenamefont
  {Khemani}, \citenamefont {Huse}, \citenamefont {Bloch},\ and\ \citenamefont
  {Gross}}]{choi2016exploring}%
  \BibitemOpen
  \bibfield  {author} {\bibinfo {author} {\bibfnamefont {J.-y.}\ \bibnamefont
  {Choi}}, \bibinfo {author} {\bibfnamefont {S.}~\bibnamefont {Hild}}, \bibinfo
  {author} {\bibfnamefont {J.}~\bibnamefont {Zeiher}}, \bibinfo {author}
  {\bibfnamefont {P.}~\bibnamefont {Schau{\ss}}}, \bibinfo {author}
  {\bibfnamefont {A.}~\bibnamefont {Rubio-Abadal}}, \bibinfo {author}
  {\bibfnamefont {T.}~\bibnamefont {Yefsah}}, \bibinfo {author} {\bibfnamefont
  {V.}~\bibnamefont {Khemani}}, \bibinfo {author} {\bibfnamefont {D.~A.}\
  \bibnamefont {Huse}}, \bibinfo {author} {\bibfnamefont {I.}~\bibnamefont
  {Bloch}}, \ and\ \bibinfo {author} {\bibfnamefont {C.}~\bibnamefont
  {Gross}},\ }\bibfield  {title} {\enquote {\bibinfo {title} {Exploring the
  many-body localization transition in two dimensions},}\ }\href
  {http://science.sciencemag.org/content/352/6293/1547} {\bibfield  {journal}
  {\bibinfo  {journal} {Science}\ }\textbf {\bibinfo {volume} {352}},\ \bibinfo
  {pages} {1547--1552} (\bibinfo {year} {2016})}\BibitemShut {NoStop}%
\bibitem [{\citenamefont {Huse}\ \emph {et~al.}(2013)\citenamefont {Huse},
  \citenamefont {Nandkishore}, \citenamefont {Oganesyan}, \citenamefont {Pal},\
  and\ \citenamefont {Sondhi}}]{huse2013localisation}%
  \BibitemOpen
  \bibfield  {author} {\bibinfo {author} {\bibfnamefont {D.~A.}\ \bibnamefont
  {Huse}}, \bibinfo {author} {\bibfnamefont {R.}~\bibnamefont {Nandkishore}},
  \bibinfo {author} {\bibfnamefont {V.}~\bibnamefont {Oganesyan}}, \bibinfo
  {author} {\bibfnamefont {A.}~\bibnamefont {Pal}}, \ and\ \bibinfo {author}
  {\bibfnamefont {S.~L.}\ \bibnamefont {Sondhi}},\ }\bibfield  {title}
  {\enquote {\bibinfo {title} {Localization-protected quantum order},}\ }\href
  {\doibase 10.1103/PhysRevB.88.014206} {\bibfield  {journal} {\bibinfo
  {journal} {Phys. Rev. B}\ }\textbf {\bibinfo {volume} {88}},\ \bibinfo
  {pages} {014206} (\bibinfo {year} {2013})}\BibitemShut {NoStop}%
\bibitem [{\citenamefont {Pekker}\ \emph {et~al.}(2014)\citenamefont {Pekker},
  \citenamefont {Refael}, \citenamefont {Altman}, \citenamefont {Demler},\ and\
  \citenamefont {Oganesyan}}]{pekker2014hilbert}%
  \BibitemOpen
  \bibfield  {author} {\bibinfo {author} {\bibfnamefont {D.}~\bibnamefont
  {Pekker}}, \bibinfo {author} {\bibfnamefont {G.}~\bibnamefont {Refael}},
  \bibinfo {author} {\bibfnamefont {E.}~\bibnamefont {Altman}}, \bibinfo
  {author} {\bibfnamefont {E.}~\bibnamefont {Demler}}, \ and\ \bibinfo {author}
  {\bibfnamefont {V.}~\bibnamefont {Oganesyan}},\ }\bibfield  {title} {\enquote
  {\bibinfo {title} {Hilbert-glass transition: New universality of
  temperature-tuned many-body dynamical quantum criticality},}\ }\href
  {\doibase 10.1103/PhysRevX.4.011052} {\bibfield  {journal} {\bibinfo
  {journal} {Phys. Rev. X}\ }\textbf {\bibinfo {volume} {4}},\ \bibinfo {pages}
  {011052} (\bibinfo {year} {2014})}\BibitemShut {NoStop}%
\bibitem [{\citenamefont {Kj\"all}\ \emph {et~al.}(2014)\citenamefont
  {Kj\"all}, \citenamefont {Bardarson},\ and\ \citenamefont
  {Pollmann}}]{kjall2014many}%
  \BibitemOpen
  \bibfield  {author} {\bibinfo {author} {\bibfnamefont {J.~A.}\ \bibnamefont
  {Kj\"all}}, \bibinfo {author} {\bibfnamefont {J.~H.}\ \bibnamefont
  {Bardarson}}, \ and\ \bibinfo {author} {\bibfnamefont {F.}~\bibnamefont
  {Pollmann}},\ }\bibfield  {title} {\enquote {\bibinfo {title} {Many-body
  localization in a disordered quantum ising chain},}\ }\href {\doibase
  10.1103/PhysRevLett.113.107204} {\bibfield  {journal} {\bibinfo  {journal}
  {Phys. Rev. Lett.}\ }\textbf {\bibinfo {volume} {113}},\ \bibinfo {pages}
  {107204} (\bibinfo {year} {2014})}\BibitemShut {NoStop}%
\bibitem [{\citenamefont {Parameswaran}\ and\ \citenamefont
  {Vasseur}(2018)}]{parameswaran2018many}%
  \BibitemOpen
  \bibfield  {author} {\bibinfo {author} {\bibfnamefont {S.~A.}\ \bibnamefont
  {Parameswaran}}\ and\ \bibinfo {author} {\bibfnamefont {R.}~\bibnamefont
  {Vasseur}},\ }\bibfield  {title} {\enquote {\bibinfo {title} {Many-body
  localization, symmetry, and topology},}\ }\href
  {https://arxiv.org/abs/1801.07731} {\bibfield  {journal} {\bibinfo  {journal}
  {arXiv:1801.07731}\ } (\bibinfo {year} {2018})}\BibitemShut {NoStop}%
\bibitem [{\citenamefont {Khemani}\ \emph {et~al.}(2016)\citenamefont
  {Khemani}, \citenamefont {Lazarides}, \citenamefont {Moessner},\ and\
  \citenamefont {Sondhi}}]{khemani2016phase}%
  \BibitemOpen
  \bibfield  {author} {\bibinfo {author} {\bibfnamefont {V.}~\bibnamefont
  {Khemani}}, \bibinfo {author} {\bibfnamefont {A.}~\bibnamefont {Lazarides}},
  \bibinfo {author} {\bibfnamefont {R.}~\bibnamefont {Moessner}}, \ and\
  \bibinfo {author} {\bibfnamefont {S.~L.}\ \bibnamefont {Sondhi}},\ }\bibfield
   {title} {\enquote {\bibinfo {title} {Phase structure of driven quantum
  systems},}\ }\href {\doibase 10.1103/PhysRevLett.116.250401} {\bibfield
  {journal} {\bibinfo  {journal} {Phys. Rev. Lett.}\ }\textbf {\bibinfo
  {volume} {116}},\ \bibinfo {pages} {250401} (\bibinfo {year}
  {2016})}\BibitemShut {NoStop}%
\bibitem [{\citenamefont {von Keyserlingk}\ \emph {et~al.}(2016)\citenamefont
  {von Keyserlingk}, \citenamefont {Khemani},\ and\ \citenamefont
  {Sondhi}}]{keyserlingk2016absolute}%
  \BibitemOpen
  \bibfield  {author} {\bibinfo {author} {\bibfnamefont {C.~W.}\ \bibnamefont
  {von Keyserlingk}}, \bibinfo {author} {\bibfnamefont {V.}~\bibnamefont
  {Khemani}}, \ and\ \bibinfo {author} {\bibfnamefont {S.~L.}\ \bibnamefont
  {Sondhi}},\ }\bibfield  {title} {\enquote {\bibinfo {title} {Absolute
  stability and spatiotemporal long-range order in {F}loquet systems},}\ }\href
  {\doibase 10.1103/PhysRevB.94.085112} {\bibfield  {journal} {\bibinfo
  {journal} {Phys. Rev. B}\ }\textbf {\bibinfo {volume} {94}},\ \bibinfo
  {pages} {085112} (\bibinfo {year} {2016})}\BibitemShut {NoStop}%
\bibitem [{\citenamefont {Moessner}\ and\ \citenamefont
  {Sondhi}(2017)}]{moessner2017equilibration}%
  \BibitemOpen
  \bibfield  {author} {\bibinfo {author} {\bibfnamefont {R.}~\bibnamefont
  {Moessner}}\ and\ \bibinfo {author} {\bibfnamefont {S.~L.}\ \bibnamefont
  {Sondhi}},\ }\bibfield  {title} {\enquote {\bibinfo {title} {Equilibration
  and order in quantum {F}loquet matter},}\ }\href
  {https://www.nature.com/nphys/journal/v13/n5/abs/nphys4106.html} {\bibfield
  {journal} {\bibinfo  {journal} {Nat. Phys.}\ }\textbf {\bibinfo {volume}
  {13}},\ \bibinfo {pages} {424} (\bibinfo {year} {2017})}\BibitemShut
  {NoStop}%
\bibitem [{\citenamefont {Zhang}\ \emph {et~al.}(2017)\citenamefont {Zhang},
  \citenamefont {Hess}, \citenamefont {Kyprianidis}, \citenamefont {Becker},
  \citenamefont {Lee}, \citenamefont {Smith}, \citenamefont {Pagano},
  \citenamefont {Potirniche}, \citenamefont {Potter}, \citenamefont
  {Vishwanath} \emph {et~al.}}]{zhang2017observation}%
  \BibitemOpen
  \bibfield  {author} {\bibinfo {author} {\bibfnamefont {J.}~\bibnamefont
  {Zhang}}, \bibinfo {author} {\bibfnamefont {P.~W.}\ \bibnamefont {Hess}},
  \bibinfo {author} {\bibfnamefont {A.}~\bibnamefont {Kyprianidis}}, \bibinfo
  {author} {\bibfnamefont {P.}~\bibnamefont {Becker}}, \bibinfo {author}
  {\bibfnamefont {A.}~\bibnamefont {Lee}}, \bibinfo {author} {\bibfnamefont
  {J.}~\bibnamefont {Smith}}, \bibinfo {author} {\bibfnamefont
  {G.}~\bibnamefont {Pagano}}, \bibinfo {author} {\bibfnamefont {I.-D.}\
  \bibnamefont {Potirniche}}, \bibinfo {author} {\bibfnamefont {A.~C.}\
  \bibnamefont {Potter}}, \bibinfo {author} {\bibfnamefont {A.}~\bibnamefont
  {Vishwanath}},  \emph {et~al.},\ }\bibfield  {title} {\enquote {\bibinfo
  {title} {Observation of a discrete time crystal},}\ }\href
  {http://www.nature.com/nature/journal/v543/n7644/full/nature21413.html}
  {\bibfield  {journal} {\bibinfo  {journal} {Nature}\ }\textbf {\bibinfo
  {volume} {543}},\ \bibinfo {pages} {217} (\bibinfo {year}
  {2017})}\BibitemShut {NoStop}%
\bibitem [{\citenamefont {Choi}\ \emph {et~al.}(2017)\citenamefont {Choi},
  \citenamefont {Choi}, \citenamefont {Landig}, \citenamefont {Kucsko},
  \citenamefont {Zhou}, \citenamefont {Isoya}, \citenamefont {Jelezko},
  \citenamefont {Onoda}, \citenamefont {Sumiya}, \citenamefont {Khemani} \emph
  {et~al.}}]{choi2017observation}%
  \BibitemOpen
  \bibfield  {author} {\bibinfo {author} {\bibfnamefont {S.}~\bibnamefont
  {Choi}}, \bibinfo {author} {\bibfnamefont {J.}~\bibnamefont {Choi}}, \bibinfo
  {author} {\bibfnamefont {R.}~\bibnamefont {Landig}}, \bibinfo {author}
  {\bibfnamefont {G.}~\bibnamefont {Kucsko}}, \bibinfo {author} {\bibfnamefont
  {H.}~\bibnamefont {Zhou}}, \bibinfo {author} {\bibfnamefont {J.}~\bibnamefont
  {Isoya}}, \bibinfo {author} {\bibfnamefont {F.}~\bibnamefont {Jelezko}},
  \bibinfo {author} {\bibfnamefont {S.}~\bibnamefont {Onoda}}, \bibinfo
  {author} {\bibfnamefont {H.}~\bibnamefont {Sumiya}}, \bibinfo {author}
  {\bibfnamefont {V.}~\bibnamefont {Khemani}},  \emph {et~al.},\ }\bibfield
  {title} {\enquote {\bibinfo {title} {Observation of discrete time-crystalline
  order in a disordered dipolar many-body system},}\ }\href
  {https://www.nature.com/nature/journal/v543/n7644/abs/nature21426.html}
  {\bibfield  {journal} {\bibinfo  {journal} {Nature}\ }\textbf {\bibinfo
  {volume} {543}},\ \bibinfo {pages} {221} (\bibinfo {year}
  {2017})}\BibitemShut {NoStop}%
\bibitem [{\citenamefont {Pal}\ \emph {et~al.}(2017)\citenamefont {Pal},
  \citenamefont {Nishad}, \citenamefont {Mahesh},\ and\ \citenamefont
  {Sreejith}}]{pal2017rigidity}%
  \BibitemOpen
  \bibfield  {author} {\bibinfo {author} {\bibfnamefont {S.}~\bibnamefont
  {Pal}}, \bibinfo {author} {\bibfnamefont {N.}~\bibnamefont {Nishad}},
  \bibinfo {author} {\bibfnamefont {T.~S.}\ \bibnamefont {Mahesh}}, \ and\
  \bibinfo {author} {\bibfnamefont {G.~J.}\ \bibnamefont {Sreejith}},\
  }\bibfield  {title} {\enquote {\bibinfo {title} {Rigidity of temporal order
  in periodically driven spins in star-shaped clusters},}\ }\href
  {https://arxiv.org/abs/1708.08443} {\bibfield  {journal} {\bibinfo  {journal}
  {arXiv:1708.08443}\ } (\bibinfo {year} {2017})}\BibitemShut {NoStop}%
\bibitem [{\citenamefont {Rovny}\ \emph {et~al.}(2018)\citenamefont {Rovny},
  \citenamefont {Blum},\ and\ \citenamefont {Barrett}}]{rovny2018observation}%
  \BibitemOpen
  \bibfield  {author} {\bibinfo {author} {\bibfnamefont {J.}~\bibnamefont
  {Rovny}}, \bibinfo {author} {\bibfnamefont {R.~L.}\ \bibnamefont {Blum}}, \
  and\ \bibinfo {author} {\bibfnamefont {S.~E.}\ \bibnamefont {Barrett}},\
  }\bibfield  {title} {\enquote {\bibinfo {title} {{Observation of discrete
  time-crystalline signatures in an ordered dipolar many-body system}},}\
  }\href {https://arxiv.org/abs/1802.00126} {\bibfield  {journal} {\bibinfo
  {journal} {arXiv:1802.00126}\ } (\bibinfo {year} {2018})}\BibitemShut
  {NoStop}%
\bibitem [{\citenamefont {Roy}\ \emph {et~al.}(2017)\citenamefont {Roy},
  \citenamefont {Lazarides}, \citenamefont {Heyl},\ and\ \citenamefont
  {Moessner}}]{roy2017dynamical}%
  \BibitemOpen
  \bibfield  {author} {\bibinfo {author} {\bibfnamefont {S.}~\bibnamefont
  {Roy}}, \bibinfo {author} {\bibfnamefont {A.}~\bibnamefont {Lazarides}},
  \bibinfo {author} {\bibfnamefont {M.}~\bibnamefont {Heyl}}, \ and\ \bibinfo
  {author} {\bibfnamefont {R.}~\bibnamefont {Moessner}},\ }\bibfield  {title}
  {\enquote {\bibinfo {title} {Dynamical potentials for non-equilibrium quantum
  many-body phases},}\ }\href {https://arxiv.org/abs/1710.09388} {\bibfield
  {journal} {\bibinfo  {journal} {arXiv:1710.09388}\ } (\bibinfo {year}
  {2017})}\BibitemShut {NoStop}%
\bibitem [{\citenamefont {Kaplan}\ \emph {et~al.}(1989)\citenamefont {Kaplan},
  \citenamefont {Horsch},\ and\ \citenamefont {Von~der
  Linden}}]{kaplan1989order}%
  \BibitemOpen
  \bibfield  {author} {\bibinfo {author} {\bibfnamefont {T.A.}\ \bibnamefont
  {Kaplan}}, \bibinfo {author} {\bibfnamefont {P.}~\bibnamefont {Horsch}}, \
  and\ \bibinfo {author} {\bibfnamefont {W.}~\bibnamefont {Von~der Linden}},\
  }\bibfield  {title} {\enquote {\bibinfo {title} {Order parameter in quantum
  antiferromagnets},}\ }\href
  {https://journals.jps.jp/doi/10.1143/JPSJ.58.3894} {\bibfield  {journal}
  {\bibinfo  {journal} {J. Phys. Soc. Jpn.}\ }\textbf {\bibinfo {volume}
  {58}},\ \bibinfo {pages} {3894--3898} (\bibinfo {year} {1989})}\BibitemShut
  {NoStop}%
\bibitem [{\citenamefont {Koma}\ and\ \citenamefont
  {Tasaki}(1993)}]{koma1993symmetry}%
  \BibitemOpen
  \bibfield  {author} {\bibinfo {author} {\bibfnamefont {T.}~\bibnamefont
  {Koma}}\ and\ \bibinfo {author} {\bibfnamefont {H.}~\bibnamefont {Tasaki}},\
  }\bibfield  {title} {\enquote {\bibinfo {title} {Symmetry breaking in
  heisenberg antiferromagnets},}\ }\href
  {https://link.springer.com/article/10.1007%2FBF02097237} {\bibfield
  {journal} {\bibinfo  {journal} {Comm. Math. Phys.}\ }\textbf {\bibinfo
  {volume} {158}},\ \bibinfo {pages} {191--214} (\bibinfo {year}
  {1993})}\BibitemShut {NoStop}%
\bibitem [{\citenamefont {D'Alessio}\ and\ \citenamefont
  {Rigol}(2014)}]{dalessio2014long}%
  \BibitemOpen
  \bibfield  {author} {\bibinfo {author} {\bibfnamefont {L.}~\bibnamefont
  {D'Alessio}}\ and\ \bibinfo {author} {\bibfnamefont {M.}~\bibnamefont
  {Rigol}},\ }\bibfield  {title} {\enquote {\bibinfo {title} {Long-time
  behavior of isolated periodically driven interacting lattice systems},}\
  }\href {\doibase 10.1103/PhysRevX.4.041048} {\bibfield  {journal} {\bibinfo
  {journal} {Phys. Rev. X}\ }\textbf {\bibinfo {volume} {4}},\ \bibinfo {pages}
  {041048} (\bibinfo {year} {2014})}\BibitemShut {NoStop}%
\bibitem [{\citenamefont {Lazarides}\ \emph {et~al.}(2014)\citenamefont
  {Lazarides}, \citenamefont {Das},\ and\ \citenamefont
  {Moessner}}]{lazarides2014equilibrium}%
  \BibitemOpen
  \bibfield  {author} {\bibinfo {author} {\bibfnamefont {A.}~\bibnamefont
  {Lazarides}}, \bibinfo {author} {\bibfnamefont {A.}~\bibnamefont {Das}}, \
  and\ \bibinfo {author} {\bibfnamefont {R.}~\bibnamefont {Moessner}},\
  }\bibfield  {title} {\enquote {\bibinfo {title} {Equilibrium states of
  generic quantum systems subject to periodic driving},}\ }\href {\doibase
  10.1103/PhysRevE.90.012110} {\bibfield  {journal} {\bibinfo  {journal} {Phys.
  Rev. E}\ }\textbf {\bibinfo {volume} {90}},\ \bibinfo {pages} {012110}
  (\bibinfo {year} {2014})}\BibitemShut {NoStop}%
\bibitem [{\citenamefont {Ponte}\ \emph
  {et~al.}(2015{\natexlab{a}})\citenamefont {Ponte}, \citenamefont {Chandran},
  \citenamefont {Papi{\'c}},\ and\ \citenamefont
  {Abanin}}]{ponte2015periodically}%
  \BibitemOpen
  \bibfield  {author} {\bibinfo {author} {\bibfnamefont {P.}~\bibnamefont
  {Ponte}}, \bibinfo {author} {\bibfnamefont {A.}~\bibnamefont {Chandran}},
  \bibinfo {author} {\bibfnamefont {Z.}~\bibnamefont {Papi{\'c}}}, \ and\
  \bibinfo {author} {\bibfnamefont {D.~A.}\ \bibnamefont {Abanin}},\ }\bibfield
   {title} {\enquote {\bibinfo {title} {Periodically driven ergodic and
  many-body localized quantum systems},}\ }\href
  {https://www.sciencedirect.com/science/article/pii/S0003491614003212}
  {\bibfield  {journal} {\bibinfo  {journal} {Annals of Physics}\ }\textbf
  {\bibinfo {volume} {353}},\ \bibinfo {pages} {196--204} (\bibinfo {year}
  {2015}{\natexlab{a}})}\BibitemShut {NoStop}%
\bibitem [{\citenamefont {Lazarides}\ \emph {et~al.}(2015)\citenamefont
  {Lazarides}, \citenamefont {Das},\ and\ \citenamefont
  {Moessner}}]{lazarides2015fate}%
  \BibitemOpen
  \bibfield  {author} {\bibinfo {author} {\bibfnamefont {A.}~\bibnamefont
  {Lazarides}}, \bibinfo {author} {\bibfnamefont {A.}~\bibnamefont {Das}}, \
  and\ \bibinfo {author} {\bibfnamefont {R.}~\bibnamefont {Moessner}},\
  }\bibfield  {title} {\enquote {\bibinfo {title} {Fate of many-body
  localization under periodic driving},}\ }\href {\doibase
  10.1103/PhysRevLett.115.030402} {\bibfield  {journal} {\bibinfo  {journal}
  {Phys. Rev. Lett.}\ }\textbf {\bibinfo {volume} {115}},\ \bibinfo {pages}
  {030402} (\bibinfo {year} {2015})}\BibitemShut {NoStop}%
\bibitem [{\citenamefont {Ponte}\ \emph
  {et~al.}(2015{\natexlab{b}})\citenamefont {Ponte}, \citenamefont
  {Papi\ifmmode~\acute{c}\else \'{c}\fi{}}, \citenamefont {Huveneers},\ and\
  \citenamefont {Abanin}}]{ponte2015many}%
  \BibitemOpen
  \bibfield  {author} {\bibinfo {author} {\bibfnamefont {P.}~\bibnamefont
  {Ponte}}, \bibinfo {author} {\bibfnamefont {Z.}~\bibnamefont
  {Papi\ifmmode~\acute{c}\else \'{c}\fi{}}}, \bibinfo {author} {\bibfnamefont
  {F.}~\bibnamefont {Huveneers}}, \ and\ \bibinfo {author} {\bibfnamefont
  {D.~A.}\ \bibnamefont {Abanin}},\ }\bibfield  {title} {\enquote {\bibinfo
  {title} {Many-body localization in periodically driven systems},}\ }\href
  {\doibase 10.1103/PhysRevLett.114.140401} {\bibfield  {journal} {\bibinfo
  {journal} {Phys. Rev. Lett.}\ }\textbf {\bibinfo {volume} {114}},\ \bibinfo
  {pages} {140401} (\bibinfo {year} {2015}{\natexlab{b}})}\BibitemShut
  {NoStop}%
\bibitem [{\citenamefont {Bordia}\ \emph {et~al.}(2017)\citenamefont {Bordia},
  \citenamefont {L{\"u}schen}, \citenamefont {Schneider}, \citenamefont
  {Knap},\ and\ \citenamefont {Bloch}}]{bordia2017periodically}%
  \BibitemOpen
  \bibfield  {author} {\bibinfo {author} {\bibfnamefont {P.}~\bibnamefont
  {Bordia}}, \bibinfo {author} {\bibfnamefont {H.}~\bibnamefont {L{\"u}schen}},
  \bibinfo {author} {\bibfnamefont {U.}~\bibnamefont {Schneider}}, \bibinfo
  {author} {\bibfnamefont {M.}~\bibnamefont {Knap}}, \ and\ \bibinfo {author}
  {\bibfnamefont {I.}~\bibnamefont {Bloch}},\ }\bibfield  {title} {\enquote
  {\bibinfo {title} {Periodically driving a many-body localized quantum
  system},}\ }\href
  {http://www.nature.com/nphys/journal/v13/n5/full/nphys4020.html} {\bibfield
  {journal} {\bibinfo  {journal} {Nat. Phys.}\ }\textbf {\bibinfo {volume}
  {13}},\ \bibinfo {pages} {460} (\bibinfo {year} {2017})}\BibitemShut
  {NoStop}%
\bibitem [{\citenamefont {Reitter}\ \emph {et~al.}(2017)\citenamefont
  {Reitter}, \citenamefont {N\"ager}, \citenamefont {Wintersperger},
  \citenamefont {Str\"ater}, \citenamefont {Bloch}, \citenamefont {Eckardt},\
  and\ \citenamefont {Schneider}}]{reitter2017interaction}%
  \BibitemOpen
  \bibfield  {author} {\bibinfo {author} {\bibfnamefont {M.}~\bibnamefont
  {Reitter}}, \bibinfo {author} {\bibfnamefont {J.}~\bibnamefont {N\"ager}},
  \bibinfo {author} {\bibfnamefont {K.}~\bibnamefont {Wintersperger}}, \bibinfo
  {author} {\bibfnamefont {C.}~\bibnamefont {Str\"ater}}, \bibinfo {author}
  {\bibfnamefont {I.}~\bibnamefont {Bloch}}, \bibinfo {author} {\bibfnamefont
  {A.}~\bibnamefont {Eckardt}}, \ and\ \bibinfo {author} {\bibfnamefont
  {U.}~\bibnamefont {Schneider}},\ }\bibfield  {title} {\enquote {\bibinfo
  {title} {Interaction dependent heating and atom loss in a periodically driven
  optical lattice},}\ }\href {\doibase 10.1103/PhysRevLett.119.200402}
  {\bibfield  {journal} {\bibinfo  {journal} {Phys. Rev. Lett.}\ }\textbf
  {\bibinfo {volume} {119}},\ \bibinfo {pages} {200402} (\bibinfo {year}
  {2017})}\BibitemShut {NoStop}%
\bibitem [{\citenamefont {Eckardt}(2017)}]{eckardt2017atomic}%
  \BibitemOpen
  \bibfield  {author} {\bibinfo {author} {\bibfnamefont {A.}~\bibnamefont
  {Eckardt}},\ }\bibfield  {title} {\enquote {\bibinfo {title} {Colloquium:
  Atomic quantum gases in periodically driven optical lattices},}\ }\href
  {\doibase 10.1103/RevModPhys.89.011004} {\bibfield  {journal} {\bibinfo
  {journal} {Rev. Mod. Phys.}\ }\textbf {\bibinfo {volume} {89}},\ \bibinfo
  {pages} {011004} (\bibinfo {year} {2017})}\BibitemShut {NoStop}%
\bibitem [{\citenamefont {Mussardo}(2010)}]{mussardo2010statistical}%
  \BibitemOpen
  \bibfield  {author} {\bibinfo {author} {\bibfnamefont {G.}~\bibnamefont
  {Mussardo}},\ }\href
  {https://global.oup.com/academic/product/statistical-field-theory-9780199547586?cc=gb&lang=en&}
  {\emph {\bibinfo {title} {Statistical field theory: {An} introduction to
  exactly solved models in statistical physics}}}\ (\bibinfo  {publisher}
  {Oxford University Press},\ \bibinfo {year} {2010})\BibitemShut {NoStop}%
\bibitem [{\citenamefont {Touchette}(2010)}]{touchette2010methods}%
  \BibitemOpen
  \bibfield  {author} {\bibinfo {author} {\bibfnamefont {H.}~\bibnamefont
  {Touchette}},\ }\bibfield  {title} {\enquote {\bibinfo {title} {Methods for
  calculating nonconcave entropies},}\ }\href
  {http://iopscience.iop.org/article/10.1088/1742-5468/2010/05/P05008/meta}
  {\bibfield  {journal} {\bibinfo  {journal} {J. Stat. Mech.}\ }\textbf
  {\bibinfo {volume} {2010}},\ \bibinfo {pages} {P05008} (\bibinfo {year}
  {2010})}\BibitemShut {NoStop}%
\bibitem [{\citenamefont {Medvedyeva}\ \emph {et~al.}(2016)\citenamefont
  {Medvedyeva}, \citenamefont {Prosen},\ and\ \citenamefont {\ifmmode
  \check{Z}\else \v{Z}\fi{}nidari\ifmmode~\check{c}\else
  \v{c}\fi{}}}]{medvedyeva2016influence}%
  \BibitemOpen
  \bibfield  {author} {\bibinfo {author} {\bibfnamefont {M.~V.}\ \bibnamefont
  {Medvedyeva}}, \bibinfo {author} {\bibfnamefont {T.}~\bibnamefont {Prosen}},
  \ and\ \bibinfo {author} {\bibfnamefont {M.}~\bibnamefont {\ifmmode
  \check{Z}\else \v{Z}\fi{}nidari\ifmmode~\check{c}\else \v{c}\fi{}}},\
  }\bibfield  {title} {\enquote {\bibinfo {title} {Influence of dephasing on
  many-body localization},}\ }\href {\doibase 10.1103/PhysRevB.93.094205}
  {\bibfield  {journal} {\bibinfo  {journal} {Phys. Rev. B}\ }\textbf {\bibinfo
  {volume} {93}},\ \bibinfo {pages} {094205} (\bibinfo {year}
  {2016})}\BibitemShut {NoStop}%
\bibitem [{\citenamefont {Fischer}\ \emph {et~al.}(2016)\citenamefont
  {Fischer}, \citenamefont {Maksymenko},\ and\ \citenamefont
  {Altman}}]{fischer2016dynamics}%
  \BibitemOpen
  \bibfield  {author} {\bibinfo {author} {\bibfnamefont {M.~H.}\ \bibnamefont
  {Fischer}}, \bibinfo {author} {\bibfnamefont {M.}~\bibnamefont {Maksymenko}},
  \ and\ \bibinfo {author} {\bibfnamefont {E.}~\bibnamefont {Altman}},\
  }\bibfield  {title} {\enquote {\bibinfo {title} {Dynamics of a
  many-body-localized system coupled to a bath},}\ }\href {\doibase
  10.1103/PhysRevLett.116.160401} {\bibfield  {journal} {\bibinfo  {journal}
  {Phys. Rev. Lett.}\ }\textbf {\bibinfo {volume} {116}},\ \bibinfo {pages}
  {160401} (\bibinfo {year} {2016})}\BibitemShut {NoStop}%
\bibitem [{\citenamefont {Levi}\ \emph {et~al.}(2016)\citenamefont {Levi},
  \citenamefont {Heyl}, \citenamefont {Lesanovsky},\ and\ \citenamefont
  {Garrahan}}]{levi2016robustness}%
  \BibitemOpen
  \bibfield  {author} {\bibinfo {author} {\bibfnamefont {E.}~\bibnamefont
  {Levi}}, \bibinfo {author} {\bibfnamefont {M.}~\bibnamefont {Heyl}}, \bibinfo
  {author} {\bibfnamefont {I.}~\bibnamefont {Lesanovsky}}, \ and\ \bibinfo
  {author} {\bibfnamefont {J.~P.}\ \bibnamefont {Garrahan}},\ }\bibfield
  {title} {\enquote {\bibinfo {title} {Robustness of many-body localization in
  the presence of dissipation},}\ }\href {\doibase
  10.1103/PhysRevLett.116.237203} {\bibfield  {journal} {\bibinfo  {journal}
  {Phys. Rev. Lett.}\ }\textbf {\bibinfo {volume} {116}},\ \bibinfo {pages}
  {237203} (\bibinfo {year} {2016})}\BibitemShut {NoStop}%
\bibitem [{\citenamefont {Lazarides}\ and\ \citenamefont
  {Moessner}(2017)}]{lazarides2017fate}%
  \BibitemOpen
  \bibfield  {author} {\bibinfo {author} {\bibfnamefont {A.}~\bibnamefont
  {Lazarides}}\ and\ \bibinfo {author} {\bibfnamefont {R.}~\bibnamefont
  {Moessner}},\ }\bibfield  {title} {\enquote {\bibinfo {title} {Fate of a
  discrete time crystal in an open system},}\ }\href {\doibase
  10.1103/PhysRevB.95.195135} {\bibfield  {journal} {\bibinfo  {journal} {Phys.
  Rev. B}\ }\textbf {\bibinfo {volume} {95}},\ \bibinfo {pages} {195135}
  (\bibinfo {year} {2017})}\BibitemShut {NoStop}%
\end{thebibliography}%
\end{document}